\documentclass[10pt, letterpaper]{IEEEtrans3}
\usepackage[latin9]{inputenc}
\usepackage{graphicx,color}
\usepackage{xspace,subfigure,epsfig,multirow}
\usepackage{xspace}
\usepackage{graphicx,color}
\usepackage{times,amsfonts,amssymb,amsmath,setspace}
\usepackage{setspace}
\usepackage{comment,url,enumerate}
\usepackage{lipsum}
\usepackage{multicol}
\usepackage[algoruled,medskip,dontprintsemicolon,linesnumbered,Algorithm]{algorithm}
\usepackage[noend]{algpseudocode}
\usepackage{float}
\usepackage{stfloats,mathtools} 
\usepackage{tkz-graph}
\usetikzlibrary{trees}

\DeclareMathAlphabet\mathbfcal{OMS}{cmsy}{b}{n}

\usepackage{bbm}\usepackage{mathrsfs}\usepackage{float}
\usepackage{cases}

\usepackage{graphicx,url,bm,subfigure,comment}
\usepackage{epsfig,textcomp}
\usepackage{times,amsfonts,amssymb,amsmath,setspace}

\newcommand{\dd}{{\rm\,d}} 
\newcommand{\1}{\mathbbm 1}

\newcommand{\Ec}{\mathcal E}
\newcommand{\Ac}{\mathcal A}
\newcommand{\Bc}{\mathcal B}
\newcommand{\Zc}{\mathcal Z}
\newcommand{\Good}{\mathcal A'}
\newcommand{\Bad}{\mathcal A''}
\newcommand{\GoodB}{\mathcal B'}
\newcommand{\BadB}{\mathcal B''}

\newtheorem{teorema}{\bf Theorem}
\newtheorem{corollario}{\bf Corollary}
\newtheorem{lemma}{\bf Lemma}

\newtheorem{proposizione}{\bf Proposition}

\def\ind{{\rm 1\hspace{-0.90ex}1}}

\newcommand{\Dc}{\mathcal D}

\newcommand{\Tc}{{\mathcal G}_{\text{T}}} 
\newcommand{\Gc}{\mathcal G}
\newcommand{\Mc}{\mathcal M}
\newcommand{\Hc}{\mathcal H}
\newcommand{\Kc}{\mathcal K}
\newcommand{\Pc}{\mathcal P}

\newcommand{\Vc}{\mathcal V}

\newcommand{\Nc}{\mathcal N}
\newcommand{\Cc}{\mathcal C}

\newcommand{\Ecb}{\mathbfcal E}
\newcommand{\Vcb} {\mathbfcal V}

\newcommand{\xb}{\mathbf x}
\newcommand{\zv}{\mathbf z}

\newcommand{\ls}[1]
   {\dimen0=\fontdimen6\the\font
    \lineskip=#1\dimen0
    \advance\lineskip.5\fontdimen5\the\font
    \advance\lineskip-\dimen0
    \lineskiplimit=.9\lineskip
    \baselineskip=\lineskip
    \advance\baselineskip\dimen0
    \normallineskip\lineskip
    \normallineskiplimit\lineskiplimit
    \normalbaselineskip\baselineskip
    \ignorespaces
}
\makeatother

\global\long\def\1{\mathbbm1}

\global\long\def\P{\mathbbm P}
\def\ind{{\rm 1\hspace{-0.90ex}1}}

\begin{document}
%

\title{Impact of Clustering on the Performance\\ of Network
  De-anonymization \thanks{C.F. Chiasserini and E. Leonardi are with Departimento di Elettronica e Telecomunicazioni, Politecnico di Torino, Italy,
email: surname@polito.it:  M. Garetto is with  Dipartimento di Informatica, 
Universita' di Torino, Italy email: michele.garetto@unito.it}
}
%
%
%
%
%

%
\author{C.F. Chiasserini, M. Garetto, E. Leonardi}

\maketitle
\begin{abstract}
Recently, graph matching algorithms have been
successfully applied to the problem of network de-anonymization,
in which nodes (users) participating to more than one social network are identified
only by means of the structure of their links to other members. This procedure 
exploits an initial set of seed nodes large enough to trigger a 
percolation process which correctly matches almost all 
other nodes across the different social networks. Our main contribution is to show the crucial role
played by clustering, which is a ubiquitous feature of realistic 
social network graphs (and many other systems). 
Clustering has both the effect of making matching algorithms
more vulnerable to errors, and the potential to dramatically
reduce the number of seeds needed to trigger percolation, thanks
to a wave-like propagation effect. We demonstrate these facts   
by considering a fairly general class of random geometric graphs 
with variable clustering level, and showing how clever algorithms 
can achieve surprisingly good performance while containing
matching errors.  
 
\end{abstract}



\section{Introduction}
The advent of online social networks, and their massive worldwide penetration,
can be well considered as one of the most influential changes brought
by information and communication technologies into our lives
during the last decade, with profound impact on all aspects of economy, society
and culture. The extraordinary capitalization of the companies running these
(typically free) online services can be explained by the huge amount of valuable 
information that can be extracted from the traces of activities
performed by billions of users. Such information allows, for example, to build
user profiles that can be effectively used for targeted advertisements,  
marketing and social surveys, and many other profitable business run by service providers 
and third parties. Privacy concerns raised by the collection, analysis and distribution
of personal data, exposed more or less consciously by active users,  
have been recently hotly debated in the media. 
User privacy is especially threatened when 
data collected 
from different systems is combined together to construct richer and 
more accurate user profiles. 

In this work we are specifically concerned with the problem  
of identifying users participating to different online social
networks\footnote{More in general, we are interested in any sort of communication system
assigning some kind of (unique) ID to users, typically 
as a result of a new registration/account creation (including
traditional communication services such as email and 
cellular networks).}. We emphasize that this problem
can be perceived by people in totally different ways. Some 
users would prefer to hide any Personal Identifiable Information (PII)
while using a service, and they see any attempt to 
correlate accounts created in different systems as a severe 
violation of their privacy. Other  users instead 
are more than happy to merge or link together their various accounts,
as this turns out to be convenient to the user itself. 
For example, 
\lq social logins'  
allow users to use existing accounts on social networks 
to directly sign into other services (different applications, websites,
public Wi-Fi hotspots). 

In our work, we are specifically interested in privacy issues, 
and consider the case of an \lq attacker' trying to identify users belonging to 
two different social networks (without their consent).       
Recently, security experts have made the dramatic discovery 
that user privacy cannot be guaranteed when traces of communication 
activities are made available after applying the simple anonymization
procedure which replaces real ID's by random labels  \cite{Narayanan}. 

A standard way to formalize the user identification problem       
is  the following: each communication system (e.g., a given social network)
generates (from the traces of user activities) a \lq contact graph'
in which nodes represent anonymized users, and edges
denote who has come in contact with whom. 
The attacker then runs a {\em graph matching} algorithm
on the contact graphs  generated by  different systems,
which in the hardest case can make use only of the topologies  
of these graphs, without any additional side information \cite{daniel}.
The majority of algorithms proposed so far to achieve this goal
are facilitated by an initial set of already matched nodes (called seeds).
This is actually a realistic case, since, as explained above,  
some users explicitly link their accounts in different systems \lq for free'.
Many proposed matching strategies, based
on heuristic algorithms, work by progressively expanding the set of already matched 
nodes, trying to identify all of the other nodes
\cite{Narayanan,peng,lattanzi}.
In particular, in their seminal paper 
Narayanan and Shmatikov \cite{Narayanan} were able to
identify a large fraction of users having account on both 
Twitter and Flickr (with only 12\% error ratio).

Significant progress has also been made towards
theoretical understanding of the feasibility of network de-anonymization 
(in the first place), and of the asymptotic performance
of graph matching algorithms applied to large systems. 
Recent analytical work has adopted the following convenient 
probabilistic generation model for two contact graphs $\Gc_1$ and $\Gc_2$: we consider the (inaccessible)
\lq ground-truth' graph $\Tc$ representing true social relationships 
among people, and then assume that $\Gc_1$ is obtained by 
independently sampling each edge of $\Tc$ with probability $s$
(similarly, and independently, $\Gc_2$).
Specifically, when the social network $\Tc$ is modeled as an Erd\"{o}s--R\'{e}nyi random graph, 
it has been shown in \cite{pedarsani} that, under mild conditions, 
users participating in two different social networks can be successfully matched by an attacker with 
unlimited computation power, even without seeds. 
In the case of Erd\"{o}s--R\'{e}nyi random graphs, in
\cite{Grossglauser} authors have also 
proposed a practical identification algorithm based on bootstrap
percolation
\cite{Janson}
and they have shown an interesting phase transition phenomenon
in the number of seeds that are required for network
de-anonymization. 
The results in \cite{Grossglauser} have been recently 
extended to the more realistic case in which contact graphs are scale-free 
(power law) random graphs. In particular, by modeling them 
as  Chung-Lu graphs, \cite{nostro-infocom} and \cite{tobias}
have independently shown that a much smaller set of seeds
is sufficient to trigger the percolation-based matching
process originally studied in Erd\"{o}s--R\'{e}nyi graphs.

While previous work has captured the impact of 
power-law degree distribution on percolation graph matching, 
another essential feature of real social networks, namely,
clustering, has not been investigated so far. 
Interestingly, in \cite{Grossglauser} authors attempted to 
apply their basic algorithm also to highly clustered 
random geometric graphs, observing almost total failure 
(error rates above 50\%). This preliminary finding
has been the starting point of our work. 
In this paper we consider a fairly general model of random geometric graphs
that allows us to incorporate various levels 
of clustering in contact graph, 
without concurrently generating a scale-free structure.
By so doing, we separate the (unkown) impact of clustering
from the (known) impact of power law degree, going back to 
the original case of Erd\"{o}s--R\'{e}nyi graphs
and exploring a totally different, \lq orthogonal' direction.
Our main findings are as follows: 

(i) Clustered networks
can be indeed largely prone to matching errors when we naively apply
the 
method proposed in
\cite{Grossglauser}.  
Such errors can be mitigated and asymptotically eliminated
by an improved matching algorithm still based on bootstrap 
percolation; 

(ii) Once errors are eliminated, clustering turns out to have 
a surprising beneficial effect on the performance of graph matching, thanks
to a wave-like propagation phenomenon that allows to
progressively identify all nodes  starting from a very small,
{\em compact} set of seeds;

(iii) In contrast with previous results derived for  Erd\"{o}s--R\'{e}nyi  and Chung-Lu graphs in
 \cite{Grossglauser,nostro-infocom}, we show that the minimum number
 of  seeds required for network de-anonymization increases 
 as  the  average node degree of the graph grows. 

Our  results are  qualitatively validated via experiments 
with real social network graphs. We emphasize that, 
although we focus  on network de-anonymization,
we do not cast our results exclusively to this problem.  
Indeed, the results we derive have much broader applicability  
since graph matching is a general problem arising in many different 
domains, ranging from computer graphics to bioinformatics.



\begin{figure}[t]
\centering
\includegraphics[width=2in]{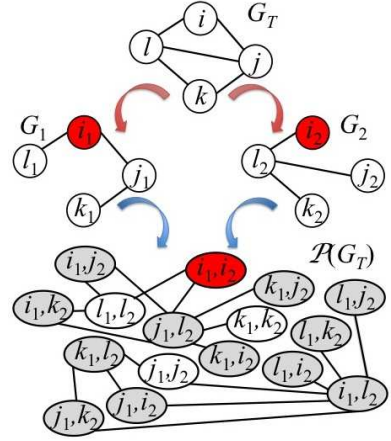}
\caption{An example of $\Gc_1$ and $\Gc_2$ obtained from $\Tc$ by
  independent edge sampling, and of the pairs graph $\Pc(\Tc)$. Seeds
are highlighted in red. In $\Pc(\Tc)$,  good pairs are
highlighted in white and bad pairs in grey.}
\label{fig:graphgeneration}
\end{figure}

\section{Notation and preliminaries\label{sec:notmodel}}


Without loss of generality, we assume that $\Tc(\Vc,\Ec)$, 
$\Gc_1(\Vc_1,\Ec_1)$ and $\Gc_2(\Vc_2,\Ec_2)$ 
have the same set of nodes (or vertices) with cardinality $n$, i.e., $\Vc_1=\Vc_2=\Vc$
\footnote{This assumption can be easily removed by considering that only the intersection of vertices
belonging to $\Gc_1$ and $\Gc_2$ has to be de-anonymized.}. 
Similarly to previous work \cite{lattanzi,pedarsani,Grossglauser,nostro-infocom,tobias}
we assume that edges in $\Gc_1$ and $\Gc_2$ are obtained by independently sampling
each edge of $\Tc$ with probability $s$. Specifically, each edge in 
$\Tc$ is assumed to be (independently) sampled twice, the first time
to determine its presence in $\Ec_1$, the second time to determine its presence
in $\Ec_2$. 
This  model is a reasonable 
approximation of real systems which permits obtaining
fundamental analytical insights. 



To match $\Gc_1$ and $\Gc_2$, we build the 
pairs graph $\Pc(\Vcb, \Ecb)$, with $\Vcb \subseteq \Vc_1\times \Vc_2$ and  
$\Ecb \subseteq \Ec_1 \times \Ec_2$. 
In  $\Pc(\Vcb, \Ecb)$ there exists an edge between $[i_1,j_2]$ and
$[k_1,l_2]$ iff edge $(i_1,k_1)\in \Ec_1$ and edge $(j_2,l_2)\in
\Ec_2$. 
We will  slightly abuse the notation and denote the pair
graph associated to a generic ground-truth graph $\Tc$
simply as $\Pc(\Tc)$. Fig. \ref{fig:graphgeneration}
shows the pairs graph built from a toy example.


We will refer to pairs  $[i_1,i_2] \in \Pc(\Tc)$, whose vertices correspond to the
same vertex $i\in \Tc$, as good pairs, and to all others (e.g.,
$[i_1,j_2]$) as bad pairs. 
Also, we will refer to two pairs such as $[i_1,j_2]$ and  $[i_1,l_2]$,
or  $[i_1,j_2]$ and  $[k_1,j_2]$, 
as conflicting.  Finally, 
two adjacent pairs on $\Pc(\Tc)$ will be referred to as neighbors.  
%
%
The seed set\footnote{We will refer to the seed set as a subset of vertices, 
or, equivalently, of good vertex pairs, that have been identified a-priori.} will be denoted by $\Ac_0(n) \subset \Vc$, with 
cardinality $a_0$.

We now briefly describe the Percolation Graph Matching (PGM)
algorithm originally proposed in \cite{Grossglauser}. 
The PGM algorithm maintains an integer counter (initialized to zero)
for any pair of $\Pc(\Tc)$ that may   still be matched.
It exploits a set $\Ac_t$, indexed by time step $t$, which 
 is initialized (for $t=0$) with the seed pairs.  
At any given time $t \geq 0$,  
 the PGM algorithm extracts  at random one pair  from $\Ac_t$  matching it, and 
 increases by one the counter associated to each of its neighbor pair
in $\Pc(\Tc)$. 
Then the algorihm adds to $\Ac_{t+1}$   all pairs whose counter 
has reached $r$ at time $t$ with the exception of 
those pairs that  are in conflict  with either  any of the already matched pairs or any of the pairs in $\Ac_t$.
The algorithms stops when $\Ac_t = \emptyset$.
It is straightforward to see that PGM takes at most $n$ steps to
terminate. 

In the case where $\Tc$ is an  
Erd\"{o}s--R\'{e}nyi random graph, previous work \cite{Grossglauser} has 
established the following lower bound on the number of seeds 
that are needed to correctly match almost all 
nodes without errors. 


\vspace{2mm}

{\bf Critical seed set size for Erd\"os-R\'enyi graphs \cite{Grossglauser}.} 
{\em Let $\Tc$ be an  Erd\"os-R\'enyi  random graph $G(m,p)$.
Let  $r\ge 4$.
Denote by $a_c$ the critical seed set size: 
\begin{equation} \label{ac}
 a_c=\left(1-\frac{1}{r}\right) \left( \frac{(r-1)!}{m(ps^2)^r}\right)^{\frac{1}{r-1}}\, .
\end{equation}
For $m^{-1} \ll ps^2 \le s^2 m^{-\frac{3.5}{r}}$, we have 
that, if  $a_o/a_c\to a>1$, the PGM algorithm matches w.h.p. a number
of good pairs 
equal to $m-o(m)$ (i.e., all vertex
pairs except for a negligible fraction) with no errors.}

\vspace{2mm}

{\bf Critical seed set size for random graphs bounded by
  Erd\"os-R\'enyi graphs.} 
Let $\Hc(\Vc,\Ec_H)$ and $\Kc(\Vc,\Ec_K)$  be two random graphs insisting on the same 
set of vertices $\Vc$, where $\Ec_H \subseteq  \Ec_K$, i.e.,  
 $\Ec_H$  can be obtained by sampling $\Ec_K$.  We  define the following  partial order relationship:
$\Hc(\Vc,\Ec_H) \le_{st} \Kc(\Vc,\Ec_K)$. 
Given that, below we extend our result in \cite{nostro-infocom}.
\begin{teorema} \label{propLU}
Consider   $\Tc$  sastisfying:
$ G(m,p_{\min})  \le_{st}  \Tc \le_{st} G(m,p_{\max})$ with $p_{\min}\leq p_{\max}$.
Applying the PGM algorithm to $\Pc(\Tc)$  guarantees that 
$m-o(m) $ good pairs are matched with no errors 
w.h.p., provided that:

\noindent
1. $m\to \infty$; 

\noindent
2. $p_{\min}=\Theta(p_{\max})$ and $p_{\min} \gg m^{-1}$;

\noindent
3. $ p_{\max} \le  m^{-\frac{3.5}{r}}$;

\noindent
4. $\lim_{m\to \infty} a_o/a_c>1$,  with $a_c$ computed  from (\ref{ac}) by   setting $p=p_{\min}$.


\noindent
Also, under conditions 1)-4),  the PGM  successfully matches w.h.p. all the  correct  pairs  (with no errors)
also in any  subgraph $\Tc'$ of $\Tc$  that comprises a finite
fraction of vertices of  $\Tc$  and all the edges between the selected vertices.
The proof can be found in Appendix~\ref{app:theorem2}
\begin{corollario} \label{propLU-1}
Under the same conditions as in Theorem~\ref{propLU}, the PGM algorithm  can be 
successfully applied  to an imperfect pairs graph
$\hat{\Pc}_x \subset \Pc(\Tc)$ comprising a finite fraction of the pairs in $\Pc(\Tc)$ and satisfying  
the following constraint: 
a bad pair $[i_1,j_2] \in \Pc(\Tc)$  is  included in $\hat{\Pc}(\Tc)$
only if either    $[i_1,i_2]$ or $[j_1,j_2]$ are  also in $\hat{\Pc}(\Tc)$.
\end{corollario} 
\end{teorema}


Under the above conditions, the objective of this work is to design
and analyze
the network de-anonymization process when the  ground-truth  
graph, $\Tc$,  exhibits different levels of nodes clustering. 
In particular, given $\Tc$,  $\Gc_1$ and $\Gc_2$, 
we aim to determine the minimum size of the
seed set  that is required to successfully 
 identify  w.h.p. all good vertex pairs in $\Pc(\Tc)$ with no errors. 
To this end, due to the big size of social network graphs, we perform an asymptotic analysis, i.e., 
we consider  the number of vertices in $\Tc$ to grow very large ($n
\to \infty$).

\section{Clustered network model}\label{sec:basic}
As detailed below, we model the social graph $\Tc$ as a geometric
random graph. 
At the end of this section, we highlight how our model well captures
node clustering and how it can represent network graphs with different values
of clustering coefficient.

We assume that nodes are located in a  $k$-dimensional 
space corresponding to the hyper-cube\footnote{To avoid border effects, we assume wrap-around conditions
(i.e., a torus topology).} ${\cal{H}} = [0,1]^k \subset
\mathbb{R}^k$, where the $k$ dimensions correspond to different
attributes of the user nodes.  
We consider the nodes to be independently and uniformly distributed
over $\Hc$.  
Given any two vertices $i,j\in \Vc$, with $i\neq j$,  
  edge $(i,j)$ exists in the graph with probability
$p_{ij}$ that depends on the Euclidean distance $d_{ij}$ 
between the respective positions of the two vertices in $\cal{H}$.
We consider the following generic law for $p_{ij}$:
\begin{equation}
p_{ij} = K(n) f(d_{ij})\,.
\label{eq:pij}
\end{equation}
In (\ref{eq:pij}),  $f$ is a non-increasing function of 
the distance, and $K(n)$ is a normalization constant introduced to impose
a desired average node degree, $D(n)$, which is assumed to 
be the same for all nodes.
It is customary in random graph models representing
realistic systems to assume that the average node degree
is not constant, but it increases with $n$ due to network 
densification. Also, although a common choice is to assume $D(n) =
\Theta(\log n)$,  
in our model we consider $D(n) =\Omega(\log n)$ so as to encompass
almost all systems of practical 
interest.

Since we are interested in the 
asymptotic performance of graph de-anonymization 
as $n$ grows large, it is convenient to further characterize
the shape of function $f$ as follows.
Let us define $C(n)$  to be at least equal to the minimal (in order sense) distance
between nodes in $\Hc$, i.e., $n^{-1/k}$.
We assume that 
$f(d)$ is equal to 1  for all distances $0 < d < C(n)$. 
This implies that $K(n)$
must be less than or equal to $1$ to obtain a proper probability function. 
For distances larger than $C(n)$, we assume that $f$ decays 
according to a power-law with exponent $\beta$, with $\beta > 0$.
In summary, 
\begin{equation}
f(d_{ij}) = \min \left\{ 1,\left( \frac{C(n)}{d_{ij}}\right)^\beta
\right\} \,. 
\end{equation}
The above characterization of the shape of $f(d)$ is fairly general
and allows accounting for different levels of node clustering.   
In particular, our random-graph model degenerates into a standard
Erd\"{o}s--R\'{e}nyi graph when 
 $C(n) = \Theta(1)$, with arbitrary $\beta$. 
For $\beta \rightarrow \infty$, instead, we have a geometric graph,
i.e., 
edges can be established only between nodes whose distance is smaller
than or equal to $C(n)$.


The average node degree is:
\begin{eqnarray}
D(n) \hspace{-3mm}&= & \hspace{-3mm}\Theta \left(n K(n) \Big(C^k(n) + C^\beta(n)\int_{C(n)}^{1}
  \rho^{k-1-\beta} \dd \rho\Big) \right) \,.\nonumber\\
\label{eq:D(n)2a}
\end{eqnarray}
Now, from  \eqref{eq:D(n)2a} it follows that 
 for  $\beta>k$  the dominant component of the  neighbors of a given node lye at a distance $\Theta(C(n))$  from it, 
while for $\beta <k$ only a marginal fraction of the neighbors of a node  lye at  distance $o(1)$ from it.
Since we are interested in graphs with  significant node 
clustering (so as to mimic real-world social networks),  we
restrict our analysis  to the case $\beta>k$. In this case,  the
average node degree is given by:
\begin{equation}
 D(n) =  \Theta(n K(n) C^k(n)).
\label{eq:D(n)2}
\end{equation}
Because by construction $K(n) \le 1$,  the average node degree 
is constrained to be $O(n C^k(n))$.
Moreover, given that  we assume $D(n)=\Omega(\log n)$, we have 
$C(n)=\Omega\left( \left(\frac{\log n}{n}\right )^{\frac{1}{k}} \right )$.

The clustering coefficient turns out to be  $\Theta(K(n))$, as a direct
consequence of the fact that the major part of the neighbors of a
 node lye at a distance $\Theta(C(n))$ from it.  In the
 following, we will slightly abuse the language and 
refer to groups of vertices that lye in sub-regions of side
$\Theta(C(n))$  as  {\em clusters}. 
Furthermore, we observe that, given the
above expressions,  
the ratio of the clustering coefficient 
($\Theta(K(n))$) to the graph density~\footnote{Given a generic graph $\Gc(\Vc,\Ec)$, the graph density is defined as  
$\frac{2|\Ec|}{|\Vc|(|\Vc|-1)}$. It can be interpreted as the probability that an edge exists between 
two randomly selected nodes of the graph.}
($\Theta(D(n)/n)$) is 
$\Theta(1/C^k(n))$. This implies that our graph exhibits a high level  
of clustering.  
 Indeed, since in general  $C^k(n)=o(1)$,  the probability that  two nodes are 
connected conditioned on the fact that they have a common
neighbor, is higher (in order sense) 
than the average probability that any two nodes are connected.   
It follows that, $K(n)$ and $C(n)$ result to be  the key model parameters through which we can directly  control
the clustering coefficient of the graph as well as  the graph density. Thus, they play a
crucial role in the analysis we present below.

\section{Overview and main results\label{sec:overview}}
In our analysis we address two cases:
 clusters with relatively sparse  structure, i.e.,   $K(n)=o((nC^k(n))^{-\gamma})$ for some  $\gamma>0$, and 
clusters with extremely dense (up to a quasi-clique)  structure,
i.e.,  $K(n)=\omega((nC^k(n)^{-\gamma}))$ for any  $\gamma>0$. 

In the  former  case  the cluster density goes to zero sufficiently
fast as the number of nodes within the cluster goes to infinity
($nC^k(n) \to \infty$). On the contrary, the latter corresponds to a cluster density
that either is bounded away from zero or goes to zero  very slowly,
with $K(n)=\Theta(1)$ being a particularly relevant  sub-case.

\begin{figure}[t.h]
\begin{center}
\begin{tikzpicture}[scale=0.65]

\draw [fill=yellow, dashed, ultra thick]  (0,0) circle (3.8 cm);
\draw [fill=orange, dashed, ultra thick]  (0,0) circle (3.0 cm);
\draw [fill=red,  dashed, ultra thick] (0,0) circle (2.2 cm);

\node[font=\scriptsize] at (0, 0) {Already de-anonimized};
\node[font=\scriptsize] at (0, 2.4) { {to be  de-anonymized }};
\node[font=\scriptsize] at (0, 3.2) { {next to be de-anonymized}};

\end{tikzpicture}
\end{center}
\vspace*{-3mm}
\caption{Graphical representation  of the de-anonymization procedure for $K(n)=o((nC^k(n))^{-\gamma})$.\label{fig:overview1}}
\end{figure}
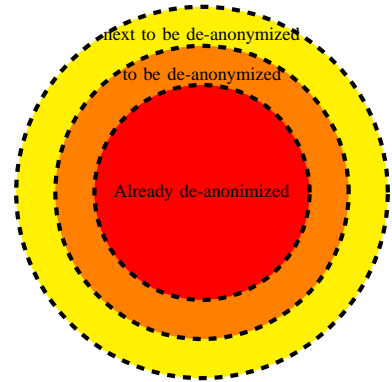

We observe that, in the case of  relatively sparse cluster structure, the density of edges
between nodes within a cluster is not excessively large  and, thus, PGM can be safely  applied without  the risk of incurring in
 significant matching errors. We therefore apply the following procedure to determine the minimum set size required for successful
graph de-anonymization. 
We assume that  the set of seeds lye in a  
small sub-region of  $\Hc$ of size  $\Theta(C(n))$ (i.e., within a
cluster). Then, through the  PGM algorithm, 
we de-anonymize all nodes that  lye  sufficiently close (within a
prefixed distance) from  the seeds. 
Once a significant bulk of pairs has been matched in this sub-region,
the de-anonymization procedure is performed by successfully matching,
at every stage,  pairs that 
are sufficiently  close to the previously matched pairs. Note
that, starting from the second stage on, we do not apply PGM any
longer but a simpler proximity-based  strategy,   matching those pairs  that have a sufficiently large number of  neighbors among the pairs matched at earlier stages.
 The way the matching procedure 
evolves is exemplified in Fig.~\ref{fig:overview1}.

\begin{table}
\label{tab:res}
\begin{center}
\caption{Main results}
\vspace{2mm}
\scriptsize
\begin{tabular}{||c||c||}
\hline
\hline
Scenario  &  Minimum seed size  \\
\hline
$K(n)=\omega((nC^k(n))^{-\gamma})$,    $\forall \gamma>0$          & $O((nC^k(n))^{\epsilon})$ $\forall \epsilon>0$ \\
\hline
$K(n)= o((nC^k(n))^{-\gamma})$, with $\gamma>0$ &
$ \Theta\left(\frac{\log nC^k(n)}{K(n)}\right)$  \\
\hline 
\end{tabular}
\end{center} 
\end{table}

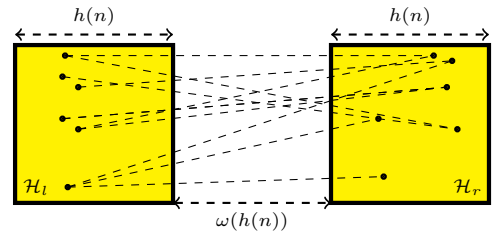
\begin{figure}[t.h]
\begin{center}
\begin{tikzpicture}[scale=0.7]
 \SetUpEdge[lw         = 1.5pt,
            color      = black,
            labelcolor = white]
  \GraphInit[vstyle=Normal] 
\draw [fill=  yellow,  ultra thick]  (0,0) rectangle (3,3);
\draw [fill= yellow,  ultra thick]  (6,0) rectangle (9,3);

\draw [fill=black,ultra thick] (1,0.3) circle (0.02 cm);
\draw [fill=black,ultra thick] (1.2,1.4) circle (0.02 cm);
\draw [fill=black,ultra thick] (0.9,1.6) circle (0.02 cm);
\draw [fill=black,ultra thick] (0.95,2.8) circle (0.02 cm);
\draw [fill=black,ultra thick] (1.2,2.2) circle (0.02 cm);
\draw [fill=black,ultra thick] (0.9,2.4) circle (0.02 cm);

\draw [fill=black,ultra thick] (7,0.5) circle (0.02 cm);
\draw [fill=black,ultra thick] (8.4,1.4) circle (0.02 cm);
\draw [fill=black,ultra thick] (6.9,1.6) circle (0.02 cm);
\draw [fill=black,ultra thick] (7.95,2.8) circle (0.02 cm);
\draw [fill=black,ultra thick] (8.2,2.2) circle (0.02 cm);
\draw [fill=black,ultra thick] (8.3,2.7) circle (0.02 cm);

\draw[dashed,thin]  (1,0.3) -- (7,0.5);
\draw[dashed,thin]  (1,0.3) -- (6.9,1.6);
\draw[dashed,thin]  (1,0.3)  -- (8.3,2.7);
\draw[dashed,thin]  (1.2,1.4) -- (7.95,2.8);
\draw[dashed,thin]  (1.2,1.4) -- (8.2,2.2);
\draw[dashed,thin]  (0.9,1.6) --  (8.2,2.2);
\draw[dashed,thin]  (0.9,1.6) -- (8.2,2.2);
\draw[dashed,thin]  (0.95,2.8) -- (7.95,2.8);
\draw[dashed,thin]  (0.95,2.8) -- (8.4,1.4);
\draw[dashed,thin]  (1.2,2.2) -- (8.3,2.7);
\draw[dashed,thin]  (0.9,2.4) -- (8.4,1.4);

\node at (0.4,0.3) { \scriptsize $\Hc_l$};
\node at (8.6,0.3) { \scriptsize  $\Hc_r$};

\draw[<->,dashed,thick]  (3,0) -- (6,0);
\draw[<->,dashed,thick]  (0,3.2) -- (3,3.2);
\draw[<->,dashed,thick]  (6,3.2) -- (9,3.2);
\node [below] at  (4.5,0)  { \scriptsize $\omega(h(n))$};
\node [above] at  (1.5,3.2)  { \scriptsize $h(n)$};
\node [above] at  (7.5,3.2)  { \scriptsize $h(n)$};
\end{tikzpicture}
\end{center}
\vspace*{-5mm}
\caption{Graphical representation  of bipartite graph construction  for $K(n)=\omega((nC^k(n)^{-\gamma}))$.\label{fig:overview2}}
\end{figure}

In the case of dense cluster structure, the whole procedure is  slightly more complex  in
light of the fact that  the clustering coefficient is larger, thus
considering short edges while running the PGM algorithm would lead to
matching a large number of bad pairs (as their counters will likely
exceed the threshold $r$). It follows that we have to ignore all 
edges  whose length is too short (shorter than a properly defined threshold  $\omega(C(n))$), in order   
 to  guarantee that almost no errors are made.   
More specifically, 
first  we consider two groups of nodes  that reside in two 
sub-regions of $\Hc$ of side $h(n)=\Theta(C(n))$,  which are taken sufficiently apart one from the
other (see Fig.~\ref{fig:overview2}). Again, we assume that an opportune  number of seeds  is included in each sub-region.  
To de-anonymize all nodes in the sub-regions, we modify the PGM algorithm so that  only  the edges between the 
two different sub-regions are exploited. Then, by leveraging the presence
of dense clusters, we  show that, given two
nodes in $\Hc$, their mutual distance can be  estimated quite precisely. 
Thus, given a sub-region where nodes have already been matched,
we can select a set of nodes that are again sufficiently apart from the
others and repeat the above procedure. The procedure can be iterated till
almost all good pairs are successfully matched. 

In  Table \ref{tab:res}, we summarize our results on the minimum size of the seed set that is required for successful network  
de-anonymization, when seeds are taken from  compact sub-regions in $\Hc$.
Observe that the minimum number of seeds  depends on both $K(n)$  and $C(n)$  while it is independent of $\beta$. 
Specifically, in the regime of  dense cluster structure (first raw of
the table), the  minimum number of seeds  can be simply  expressed in
terms of the average number of nodes  falling within a cluster ($nC^k(n)$). Indeed, a seed set whose size is equal  
to  $(nC^k(n))^\epsilon$,  for some $\epsilon$,   is enough to guarantee an almost complete successful  
network de-anonymization. In the  relevant case  in which
$C(n)=\Theta(\frac{\log n}{n})$  (i.e., when the average degree of the
graph $D(n)=\Theta(\log n)$),    the above expression degenerates into
$(\log n)^{\epsilon}$.  This last expression permits grasping immediately
the  potential impact of node clustering on de-anonymization 
techniques.  Furthermore, somehow surprisingly, the minimum seed set size increases 
 when  we increase  the  average degree of the graph nodes, by
 increasing $C(n)$. We remark that   
this is in sharp contrast with previous results derived for  Erd\"{o}s--R\'{e}nyi  and Chung-Lu graphs in
 \cite{Grossglauser,nostro-infocom}. The intuition behind this result
 is that, by increasing $C(n)$,   we
 increase also the cluster size making the problem of identifying
 nodes (users) within a cluster intrinsically more challenging.
At last, when clusters become sparser (second raw of the table), de-anonymization techniques become less effective, and the minimum seed set size shows   inverse proportional dependency on $K(n)$.




\section{Sparse clusters\label{sparse}}

In this case, we assume $K(n)=o\left((nC^k(n))^{-\gamma}\right)$, for some
$\gamma>0$, and   a set of seeds
$\Ac_0$ ($|\Ac_0|=a_0$)  
whose maximum mutual  distance is 
$d_s = O(C(n))$. 

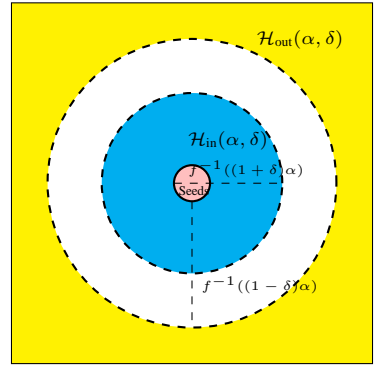
\begin{figure}[t.h]
\begin{center}
\begin{tikzpicture}[scale=0.48]
 \SetUpEdge[lw         = 1.5pt,
            color      = black,
            labelcolor = white]
  \GraphInit[vstyle=Normal] 
\draw [fill=yellow]  (-5,-5) rectangle (5,5);
\draw [fill=white,dashed, thick] (0,0) circle (4 cm);
\draw [fill= cyan, dashed, thick]  (0,0) circle (2.5 cm);
\draw [fill= pink,  thick] (0,0) circle (0.5 cm);

\draw [thin, dashed] (-0.5,0) -> (2.5,0);
\node [above] at (1.5, -0.1) {\tiny $f^{-1}((1+\delta)\alpha)$};
\node at (0, -0.2) {\tiny { Seeds}};
\node at (1, 1.2) { \scriptsize$\Hc_{\text{in}}(\alpha,\delta)$};
\node at (3, 4) {\scriptsize $\Hc_{\text{out}}(\alpha,\delta)$};
\draw [thin, dashed] (0,-0.5) -> (0,-4);
\node [right] at (0,-2.8) {\tiny $f^{-1}((1-\delta)\alpha)$};
\end{tikzpicture}
\end{center}
\vspace*{-3mm}
\caption{Graphical representation  of $\Hc_{\text{in}}(\alpha,\delta)$  and  $\Hc_{\text{out}}(\alpha,\delta)$.\label{figHin}}
\end{figure}

As the first step,  we show how nodes in $\Hc$  that lye sufficiently
close to  seeds can be identified. 
To this end, we start by defining two sub-regions, $\Hc_{\text{in}} \subset \Hc$ and
$\Hc_{\text{out}} \subset \Hc$.
Intuitively, 
$\Hc_{\text{in}}$ ($\Hc_{\text{out}}$) can be seen as a set of
points whose distance from any seed vertex is 
higher (lower) than a given threshold. More formally, denote by $\xb$ a generic point in $\Hc$ and by $\xb_\sigma$ the position 
in $\Hc$ of a generic seed vertex $\sigma$.
Then, given two positive constants $\alpha$ and $\delta$,
s.t. $\delta \le 1$ and $\alpha(1+\delta)\le 1$, we have:
\begin{align}
& \Hc_{\text{in}}(\alpha,\delta)=\Big\{ \xb \; \mbox{s.t.}  \max_{\sigma \in \Ac_0} \| \xb-\xb_\sigma \|  
\le  f^{-1}((1+ \delta)\alpha )  \Big\} \nonumber 
\end{align}
\begin{align}
&\Hc_{\text{out}}(\alpha,\delta,)=\Big\{ \xb \; \mbox{s.t.} \min_{\sigma\in \Ac_0}  
 \| \xb-\xb_\sigma \| 
 >  f^{-1}((1- \delta)\alpha )\Big\} \nonumber 
\end{align}
where $f$ is the non-increasing function defined in Section \ref{sec:basic}.  The two
sub-regions are depicted in Fig.~\ref{figHin}. Recall that, by
construction, $|\Hc_{\text{in}}|=\Theta(C^k(n))$ since 
$f(d)$ vanishes for $d \gg C(n)$.

The  theorem below proves  that, given graph $\Gc_1$ ($\Gc_2$), 
it is possible  to correctly distinguish nodes in 
$\Hc_{\text{in}}(\alpha,\delta)$ from nodes in
$\Hc_{\text{out}}(\delta, \alpha)$  by counting the number of their neighbor seeds.  
\begin{teorema} \label{teo:separation}
Given a node $i \in \Gc_1$  ($i \in \Gc_2$),  let   $S_i$ be the
number of seeds that are neighbors of $i$  on  $\Gc_1$  ($\Gc_2$). We 
say that node $i$ is 
accepted if
$S_i>\alpha s K(n) a_0$. If  $d_s=O(C(n))$ and  $a_0 =
\Omega\left(\frac{\log(nC^k(n))}{K(n)}\right)$,    then  
for an  arbitrary  $\delta>0$, the above procedure correctly accepts all 
nodes located in
$\Hc_{\text{in}}(\alpha,\delta)$ while it  
excludes all nodes located in $\Hc_{\text{out}}(\alpha,\delta)$. 
\end{teorema}
\IEEEproof See Appendix \ref{appA}.
\endIEEEproof
Note that, 
in the above statement,  $s K(n)$ is the probability that a node in $\Gc_1$  ($\Gc_2$) \footnote{Recall that $\Gc_1$  ($\Gc_2$) is a subgraph obtained from $\Tc$
by sampling the edges with probability $s$.} is connected with a seed node if their distance is 
$C(n)$ or less.  
Thus, $\alpha sK(n)  a_0$ is a threshold on  the number of  connections between a node and the $a_0$
seed vertices.   

Next, we denote by $\Nc^1(\alpha)$ and $\Nc^2(\alpha)$,  respectively, the
set of nodes from $\Gc_1$ and $\Gc_2$ that are classified as located in  $\Hc_{\text{in}}(\alpha,\delta)$.
By construction, we  have  $|\Nc^1(\alpha)|=\Theta(nC^k(n))$ and  $|\Nc^2(\alpha)|=\Theta(nC^k(n))$.
We  build 
the pairs graph $\Pc(\Nc)$ that is
induced by  the nodes of $\Gc_1$ and $\Gc_2$ that belong to, respectively, $\Nc^1(\alpha)$ and $\Nc^2(\alpha)$.
While doing this, we make sure  that a bad pair $[i_1,j_2]$ is included in  $\Pc(\Nc) $   only if either    $[i_1,i_2]$ or $[j_1,j_2]$ are  also included  
in $\Pc(\Nc)$.
This is accomplished as follows. We apply the  previous classification 
procedure twice,   using two different values $\alpha_1$ and $\alpha_2$,  with $\alpha_1>\alpha_2$, chosen in such a way that 
$\Hc_{\text{out}}(\alpha_1,\delta) \subseteq
\Hc_{\text{in}}(\alpha_2,\delta)$. 
Then we insert in 
 $\Pc(\Nc)$ all  pairs  whose constituent nodes have been selected by at least one of the  classification procedures,  
 adding  the  constraint that at least one of the nodes must have been  selected by both.
 Since by construction, no good pair $[i_1,i_2]$ exists s.t. $i_1$ falls in 
 $\Hc_{\text{in}}(\alpha_1,\delta)$ and $i_2$ in
 $\Hc_{\text{out}}(\alpha_2,\delta)$ (or viceversa), the above condition is ensured.

We then 
apply the PGM algorithm  on $\Pc(\Nc)$. Our goal is now to verify that
the  conditions
 in Theorem \ref{propLU} hold so that, applying the theorem and
 Corollary \ref{propLU-1}, we can  claim that all  good pairs in
 $\Pc(\Nc)$ 
can be  matched with no error. 
 To this end, let us define $m =\Theta(nC^k(n))$, which in order sense
 equals the number of nodes in $\Nc^1(\alpha)$ and
 $\Nc^2(\alpha)$. Then 
recall that $p_{\min}=\Theta(p_{\max})$, 
 $p_{\max}=K(n)$ and $K(n)= o(m^{-\gamma})$. Thus, for a
 sufficiently large $r$, $p_{\max} \ll
 m^{-\frac{3.5}{r}}$. Furthermore,  since by assumption $nC^k(n)K(n)=\Omega(\log n)$,
 it follows $p_{\min}\gg m^{-1}$.  At last, it is
 easy to see that $\ a_o/a_c\to \infty$. Indeed, from
 (\ref{ac}), 
 $a_c=O(1/K(n))$  while,  by assumption (see Theorem
\ref{teo:separation}), $a_0=
\Omega\left(\frac{\log(nC^k(n))}{K(n)}\right)$. 
In conclusion, we have that 
all good pairs whose nodes fall in  $\Hc_{\text{in}}(\alpha_1,\delta)$
can be correctly  matched. 

To further expand  the set of identified pairs, we can pursuit the
following simple approach. Starting from the bulk of pairs already
matched, which act as seeds, we consider a larger region that includes the previous one. By
properly setting a  threshold $r$, we  match all the pairs that have at least $r$ 
neighbors among the seeds. So doing, we successfully match w.h.p. all
good pairs in the region with no errors.  
More formally, the following theorem  allows us to claim that our approach can be successfully employed. 
\begin{teorema} \label{teo-prop}
Consider a circular region centered in 0 and of radius $\rho$, $\Dc(0,\rho)$, with $\rho \ge C(n)$. 
Given that all (or almost all) nodes    lying
within  $\Dc(0,\rho)$   have been correctly identified,  it is possible  to correctly identify  
(almost) all nodes in $\Dc(0, \rho_1) \setminus \Dc(0, \rho)$ with a probability $1- o(n^{-1})$
for $\rho_1  = \rho + C(n)/2$ when  $K(n)=o((nC^k(n))^{-\gamma})$ for some $\gamma>0$.
In addition, none of the bad pairs  
formed by nodes 
in $\Hc -  \Dc(0, \rho)$ will be identified with a probability $1- o(n^{-1})$.  This is done 
 by setting a threshold $r= \frac{n}{2}  | \Dc(0, \rho)\cap \Dc(\xb,
 C(n))|\frac{K(n)}{2}$,  with $|\xb |=\rho_1$ and identifying as good
 pairs those in $\Hc \setminus \Dc(0, \rho)$  that have at least $r$ neighbors among good  pairs in 
$\Dc(0, \rho)$. 
\end{teorema}
The proof is based on the application of standard
concentration results, namely, Chernoff bound and
inequalities in Appendix~\ref{penrose}.
The detailed proof is given in Appendix~\ref{app:teo-prop}.
Almost all good pairs can be matched w.h.p. by  iterating the matching  procedure of
Theorem \ref{teo-prop} a number of   $\Theta(1/C(n))$ times. Indeed, each time
the PGM algorithm successfully matches  
 all good  pairs whose constituent nodes  lye within a distance
 $C(n)/2$ from the bulk of previously matched pairs.
Note that Theorem \ref{teo-prop} also guarantees that jointly over all rounds  no bad pair is matched w.h.p.

\section{Dense clusters\label{sec:dense}}

The case $K(n)=\omega((nC^k(n)^{-\gamma}))$, for any  $\gamma>0$, is
significantly different from the previous one since the de-anonymization algorithm must disregard
all edges  whose length  is too short (shorter than a properly defined threshold $\omega(C(n))$) so as
to avoid errors (i.e., matching bad pairs). 
The approach we propose to address this case relies on  some results
that we initially obtain for the special case in which  $\Tc$  
is a bipartite graph.
Then we extend such results to our clustered social
graph and derive the minimum seed set size
that is required for 
graph  de-anonymization. 

\subsection{Results on bipartite graphs}
Here we   restrict our analysis to a ground-truth graph 
$\Tc$  that is an $m_l  \times m_r$  bipartite graph.   Let $\Mc_l$  denote the set of  vertices
on the left hand side (LHS), with $|\Mc_l |=m_l$, and  $\Mc_r$  the
set of vertices on the right hand side (RHS), with $|\Mc_r |=m_r$. 
We assume that  for any pair of vertices $i \in \Mc_l$ and $j \in \Mc_r$  an edge $(i,j)$ exists in the graph with 
probability $p_{ij}$, with $p_{\min} \le p_{ij}\le p_{\max}$ and $p_{\max}= \eta p_{\min}$ for some finite positive $\eta$. 
The goal here
is to  identify a minimum number of seeds $a_0$,  
 with $a_0=|\Ac_0^l|$ in $\Mc_l$  and $a_0=|\Ac_0^r|$ in $\Mc_r$,  such that
 vertices in $\Mc_l$  and  $\Mc_r$  can be correctly matched.

Let us first consider the case where $m_l=m_r=m$, 
for which the theorem below holds.
\begin{teorema} \label{maintheo}
Assume that $\Tc$ is an $m \times m$   bipartite graph and  
that  two sets of seeds, $\Ac_0^l$ and  $\Ac_0^r$, of cardinality
$a_0$,   
are available on, respectively, the LHS and  the RHS of the graph. Then 
the PGM algorithm  with threshold $r\ge 4$ correctly identifies $m-
o(m)$ good pairs w.h.p. on the RHS and the LHS of graph $\Pc(\Tc)$, 
with no errors,  
if:

\noindent
1. $m^{-1} \ll p_{\min}\le p_{\max}\le m^{-\frac{3.5 }{r}}$

\noindent
2. $\liminf_{m \to \infty} a_0/a_c>1$

\noindent
\[\mbox{where}\quad a_c= \Big( 1  - \frac{1}{r} \Big) \left( \frac{(r-1)!}{m (p_{\min} s^2)^r }   \right)^{\frac{1}{r-1}}   \,.\]
\end{teorema}
\IEEEproof See Appendix \ref{appB}.
\endIEEEproof
Theorem \ref{maintheo} can be extended to the more general case
where $m_l\neq m_r$, as shown by the corollary below.
\begin{corollario} \label{main-coro}
Assume that $\Tc$  is an $m_l \times m_r$  bipartite
graph and define $m=\min(m_l,m_r)$. Under the same assumptions of Theorem \ref{maintheo}, 
the PGM algorithm  with threshold $r\ge 4$ successfully identifies
w.h.p. $m- o(m)$ 
good pairs on both  the LHS and the RHS of $\Pc(\Tc)$,  with no errors. 
Furthermore, the PGM algorithm can be successfully  applied 
 to an imperfect pairs graph 
$\hat{\Pc}(\Tc)\subset \Pc(\Tc)$ comprising a finite fraction of pairs
on both the LHS and the RHS 
of $\Pc(\Tc)$ and satisfying  
the following constraint: 
a bad pair $[i_1,j_2] \in \Pc(\Tc) $  is  included in $\hat{\Pc}(\Tc)$
only if either    $[i_1,i_2]$ or $[j_1,j_2]$ are  also in $\hat{\Pc}(\Tc)$.
\end{corollario}
\IEEEproof 
The assertion can be proved by following the same
arguments as in Theorem \ref{maintheo} and applying Corollary \ref{propLU-1}.
\endIEEEproof
Finally, we prove the following result, which shows that all good
pairs can be matched with no errors w.h.p. 
\begin{teorema} \label{secondteo}
Consider that $\Tc$  is an $m_l \times m_r$  bipartite graph
with $m_l=\omega(\sqrt{m_r})$ and that a seed set $\Ac_0^l$ is available  on the
LHS of the graph, with
$|A_0^l|=a_0=\Theta(m_l)$.  
With probability larger than $1-e^{-\frac{m_l}{\sqrt{m_r}}}$, all the $m_r$  good pairs 
on the RHS can be successfully identified with no errors, provided that:

\noindent
1. $\frac{1}{\sqrt{m_r}} \ll p_{\min}\le  p_{\max} \ll 1$

\noindent 
2. $p_{\min}=\Theta( p_{\max} )$

\noindent
3. a matching algorithm is used on $\Pc(\Tc)$ that matches all pairs on the RHS that have at least 
$r$ adjacent  seeds on the LHS, with $r= a_0 \frac{p_{\min}}{2}$.

\noindent
The same result  holds  in case of imperfect pairs graph 
 comprising a finite fraction of all possible pairs on the RHS.
\end{teorema}
\IEEEproof 
Without loss of generality, we assume  $a_0\ge cm_r$ for some $c>0$.
The proof is obtained by applying the inequalities reported in
Appendix~\ref{penrose} 
First, observe that, given a good pair $[j_1,j_2]$ on the RHS of the
pairs graph, its number of adjacent seeds on the LHS is
$E[N_g]\ge a_0 p_{\min}=2r$. Thus, by applying inequality  
\eqref{sotto}  
and union bound, we have:
\begin{align}
P(\text{all good pairs on the RHS have at least $r$ adjacent seeds}) \nonumber\\
\ge 1 - m_r e^{-cm_l{p_{\min}} H(\frac{1}{2})}\ge 1-e^{-\frac{m_l}{\sqrt{m_r}}}\nonumber  
\end{align}
which imply that all good pairs on the RHS are successfully matched
since $m_l=\omega(\sqrt{m_r})$. 
Similarly, considering a bad pair $[j_1,k_2]$ on the RHS, its number
of adjacent seeds on the LHS is
$E[N_b]\le c m_r (p_{\max})^2\ll r$. Thus, by applying inequality  
\eqref{soprasopra} 
and union bound, we have:
\begin{align}
P(\text{all bad pairs on the RHS have less than  $r$ adjacent seeds}) \nonumber\\
\ge 1 - m_r^2 e^{-cm_l\frac{p_{\min}}{4}\log\left(\frac{p_{\min}}{ (p_{\max})^2 }\right) }\ge 1-e^{-\frac{m_l}{\sqrt{m_r}}}\,.\nonumber  
\end{align}
\endIEEEproof

\subsection{The de-anonymization procedure}
We now outline how  our proposed  de-anonymazion technique works.
First, we consider two  hyper-cubic  regions,   $\Hc_l \subset \Hc$  and
$\Hc_r \subset \Hc$,  
whose side is $h(n)=\Omega(C(n))$ and whose  distance 
is  $g(n)=\omega(C(n))$ (see Fig.~\ref{fig:overview2}). 
 Note that by construction, given two vertices $i\in \Hc_l$  and $j\in \Hc_r$,
$p_{\min}= K(n)f(g(n)+ \sqrt{k} h(n))  \le p_{ij}\le K(n)f(g(n))=p_{\max}$. 
Let us assume $p_{\max} = \eta p_{\min}$ for some constant $\eta>1$.


We then extract vertices  in
$\Hc_l$  and $\Hc_r$ from the rest of vertices so that we can  
focus on  the bipartite graph  induced by the  nodes in the two
sub-regions, along with  the edges between them.
To this end, we assume that two sufficiently large sets of seeds  are available in
$\Hc_l$  and $\Hc_r$ so that  Theorem \ref{teo:separation} can be
applied.  In this regard, observe that
we can use the same procedure  as in Section~\ref{sparse}, to make sure
that a bad pair $[i_1,j_2]$ is included in the pair graph 
only if either    $[i_1,i_2]$ or $[j_1,j_2]$ are  also included  in it. We can then  apply 
 Corollary \ref{main-coro}.


It follows that the execution of the PGM algorithm ensures  that
almost all of the  
good pairs in either the LHS  or the RHS of the pairs graph  are correctly de-anonymized. Without lack of generality, we assume that 
almost all pairs on  LHS are de-anonymized, i.e., $m_l<m_r$, and that 
a non-negligible fraction of the good pairs on the RHS
have still to be identified. Then the rest of 
good pairs  on the RHS can be matched by applying 
Theorem \ref{secondteo}.   

To further extend the de-anonymization procedure, we first observe
that it is possible to estimate in order sense the length of the edges
between two nodes, again, by exploiting the dense structure of the
clusters. 
\begin{proposizione} \label{propdist1}
Given two nodes in region $\Hc$, it is possible to estimate with arbitrary precision
their
mutual distance $d$ as far as $d \ll C(n) \left ( n K^2(n) C^k(n)\right )^\frac{1}{\beta}$.
\end{proposizione} 

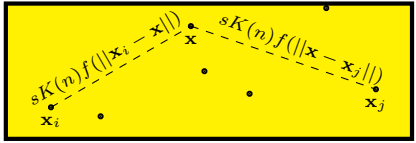
\begin{figure}[h]
\begin{center}
\begin{tikzpicture}[scale=0.6]
 \SetUpEdge[lw         = 1.5pt,
            color      = black,
            labelcolor = white]
  \GraphInit[vstyle=Normal] 

\draw [fill=  yellow,  ultra thick]  (0,0) rectangle (9,3);

\draw [fill=black,ultra thick] (1,0.7) circle (0.02 cm);
\draw [fill=black,ultra thick] (8.2,1.1) circle (0.02 cm);
\draw [fill=black,ultra thick] (4.1,2.5) circle (0.02 cm);

\draw [fill=black,ultra thick] (4.4,1.5) circle (0.02 cm);
\draw [fill=black,ultra thick] (2.1,0.5) circle (0.02 cm);

\draw [fill=black,ultra thick] (5.4,1.0) circle (0.02 cm);
\draw [fill=black,ultra thick] (7.1,2.9) circle (0.02 cm);

\node[below] at (1,0.7) {\scriptsize $\xb_i$};
\node[below] at (8.2,1.1) {\scriptsize $\xb_j$};
\node[below] at (4.1,2.5)  {\scriptsize $\xb$};

\draw[dashed,thin]  (1,0.7)  -- (4.1,2.5);
\draw[dashed,thin]  (4.1,2.5) -- (8.2,1.1);

\node[label={[rotate=30]above:{\scriptsize $sK(n)f(||\xb_i-\xb||)$}}] at  (2.4,1.2)  {};
\node[label={[rotate=-20]above:{\scriptsize $sK(n)f(||\xb-\xb_j||)$}}] at  (6.4,1.4)  {};


\end{tikzpicture}
\end{center}
\vspace*{-3mm}
\caption{Computation of $\mathbb{E}[N_{ij}]$.}
\end{figure}

\IEEEproof
Let us consider two nodes $i$ and $j$ on $\Gc_1$ ($\Gc_2$) whose mutual distance is $d_{ij}$.
Let $N_{ij}$ be the variable that represents the number of their
common neighbors. 
By construction, we have:
\begin{eqnarray}
\mathbb{E}[N_{ij}]
\hspace{-3mm}&=& \hspace{-3mm} (n-2)s^2K^2(n)\int_{\Hc} f(||\xb-\xb_i||)f(||\xb-\xb_j||){\rm d}\xb \nonumber\\
\hspace{-3mm}&=& \hspace{-3mm} \Theta(n C^k(n) K^2(n)f(d_{ij})) \,.\nonumber
\end{eqnarray}
Observe that $\mathbb{E}[N_{ij}]$ is continuous and strictly decreasing with $d_{ij}$, and thus invertible.
Now, applying Chernoff bound we can show that for any $ 0< \delta<1$ 
\[
\P\left(\frac{|N_{ij}-\mathbb{E}[N_{ij}]|}{\mathbb{E}[N_{ij}]}>\delta\right)\le e^{-c(\delta)\mathbb{E}[N_{ij}]}
\]
for a proper  constant $c(\delta)>0$.   
Furthermore for $\delta >1$
\[
\P\left(\frac{N_{ij}}{\mathbb{E}[N_{ij}]}>\delta\right)\le e^{-c(\delta)(\delta\mathbb{E}[N_{ij}] \log \delta)}
\]
Since $\mathbb{E}[N_{ij}]\to \infty $ as long as 
$d \ll C(n) \left ( n K^2(n) C^k(n)\right )^\frac{1}{\beta}$,  the assertion follows.
\endIEEEproof

We can therefore use the number of common neighbors  between  two endpoint
nodes as an estimator for their distance. 
We then set two thresholds, $d_L=\Theta(C(n)\log(n^{1/k} C(n)))$ and
$d_H=\lambda d_L$ (with $\lambda>1$), 
and we leverage the above result to correctly classify the edges departing from 
previously matched nodes into three categories: edges that are
shorter than $d_L$, edges that are longer than $d_H$ and edges of length  comprised between $d_L$ and $d_H$.
In particular, we are interested in the latter, for which the
following result holds.
\begin{proposizione}\label{prop:selectedges}
Assume $K(n)=\omega((nC^k(n))^{-\gamma})$ $\forall \gamma>0$.
Consider a set comprising a finite fraction of the nodes  of $\Gc_1$ ($\Gc_2$) that lye in a region of side $\Theta(C(n))$,
and the edges  incident to  them.
For an arbitrarily  selected $\delta>0$, w.h.p (i.e., with a probability larger than  $1- [C(n)]^{k}$) we can select all  edges whose length $d$ is 
$(1+\delta) d_L\le d\le (1-\delta)d_H$. Furthermore, no edges  whose length $d<(1-\delta)d_L$ and $d>(1+\delta)d_H$ are selected.
\end{proposizione} 

\IEEEproof
The proof follows the same scheme of proof of Theorem~\ref{teo:separation}, here we provide just a sketch.

Fix a  $\delta>0$, first we consider all edges whose  lenght does not exceed $(1-\delta)d_L$
By applying Proposition \ref{propdist1} and the union bound,
the probability that they are jointly not selected  can be bounded by:
\begin{eqnarray}
 P(\text{some edge with length  $d<(1-\delta)d_L$ is selected}) \nonumber\\
 \le  N_e e^{-c'  n C^k(n) K^2(n)f((1-\delta)d_{L})} \nonumber
\end{eqnarray}
where $N_e$ is the number of edges with length $d<(1-\delta)d_L$ and $c'$ is an opportune constant.
Now since by construction $N_e=O(nC^k(n))D(n))=O((nC^k(n))^2 K(n))$ and $d_L=\Theta(C(n)\log C(n))$ 
none of those edges is included.
Similarly we can show that all edges whose length is $(1+\delta)s_L\le d\le (1-\delta)$ are selected.

To show that none of  the edges whose length is exceeding $d_H(1+\delta)$ are selected we resort on the same ideas 
of the proof of Theorem \ref{teo:separation}. In particular,
 we   partition  such edges  into smaller  groups containing only those edges of similar lenght.  For each of  groups 
 we  have defined,
 we exploit Chernoff inequality along with the  union bound (similarly as before)  to provide an upper  bound to 
 the probability that at least one of such edges is selected. We can conclude our proof showing that previous property holds uniformly on all the groups.
\endIEEEproof

At this point, we consider a bipartite graph whose LHS is still
represented by $\Hc_l$, and whose RHS is given by the nodes that are
connected with those in  $\Hc_l$ through edges of length comprised
between $d_L$ and $d_H$. We can therefore apply Theorem
\ref{secondteo} and match w.h.p. all good pairs on the RHS, with no
errors.
The procedure is then iterated so as to successfully de-anonymize the
whole network graph. Note that, at every step 
we apply the following proposition to extract a group of matched nodes
whose mutual distance is $\Theta(C(n)))$. 
\begin{proposizione}
Assume $K(n)=\omega((nC^k(n))^{-\gamma})$ $\forall \gamma>0$.
Given a node $i$, we can set a threshold $d_T=\Theta(C(n))$ and   select all nodes in $\Gc_1$ ($\Gc_2$)
whose  estimated distance from $i$ is less than  $d_T$.
So doing, for an arbitrarily  selected $\delta>0$,  we successfully select   with a probability larger than $1-[C(n)]^{k}$ 
all  nodes whose real distance  is 
$ d\le (1-\delta)d_T$. Furthermore, no nodes  whose distance from $i$ is  $d>(1+\delta)d_T$ are selected by our algorithm.
\end{proposizione}

\IEEEproof
 The proof follows exactly te same lines as the proof of Propostion \ref{prop:selectedges}.
\endIEEEproof

\subsection{Minimum seed set size}

To explicitly derive the minimum size of the seed set, we need to
further specify   
$h(n)$ and $g(n)$, which are to be carefully selected  so as to minimize the resulting critical size 
$a_c$  in Theorem \ref{maintheo} and Corollary \ref{main-coro}.

Starting from the result provided by Theorem \ref{maintheo}, 
$a_c$ can be written as:
\begin{eqnarray} \label{seedset}
a_c &= &
\left(1-\frac{1}{r}\right) \left(
  \frac{(r-1)!}{m(p_{\min}s^2)^r}\right)^{\frac{1}{r-1}}\nonumber\\
&\le &  \left( \frac{r-1}{(m(p_{\min}s^2)^{\frac{1}{r-1}}p_{\min}s^2}\right)
  \le \frac {r-1}{p_{\min}s^2} \,.
\end{eqnarray} 
The above expression can be minimized by maximizing  $p_{\min}$, i.e.,
by minimizing $g(n)$ (recall that $p_{\min}= K(n)f(g(n)+ \sqrt{k} h(n)))$.
However,    $g(n)$ and $h(n)$ must also be  selected 
in such a way that condition 1) of Theorem \ref{maintheo} is met.
Additionally, as mentioned,  it must be ensured that 
$h(n)=\Omega(C(n))$. 
At last,  by standard concentration results, 
  $m_l$ and $m_r$  turn out to be both  $\Theta(nh^k(n))$   provided
  that  $h(n)\ge (\log n/n)^{1/k}$. 

Previous considerations induce us to fix $h(n)=\Theta(C(n))\ge
(\log n/n)^{1/k}$ (i.e., the minimum possible value in order sense),  
which corresponds to have  
 $m=\Theta(nC^k(n))$  (recall that $m=\min(m_l, m_r)$). 
We then derive $g(n)$ by forcing
 $p_{max} \approx  m^{-\frac{\alpha}{r}}$,  with $3.5 < \alpha<4$ and
 $r\ge 4$.  
Note that condition 1) of Theorem \ref{maintheo} is met since $p_{\max} $ and  $p_{\min}$ are both $\Theta(m^{-\frac{\alpha}{r}})$. 
Hence, we  have 
$p_{\max}=\Theta((n C^k(n))^{-\frac{\alpha}{r}}) $ and 
$g(n)=\Theta(n ^{\frac{\alpha}{\beta r}} [C(n)]^{1+\frac{\alpha k}{\beta r}}[K(n)]^{\frac{1}{\beta}}$)).

Given the above expression for $p_{max}$, considering that
$p_{max}=\eta p_{min}$ and using (\ref{seedset}), the minimum seed set
size can be made as small as 
\[a_c=O([n C^k(n)]^\epsilon)\] 
for any $\epsilon>0$,  by choosing $r > \frac{4}{\epsilon}$.

Finally, we remark that the obtained $a_c$ is in order sense greater than the minimum number of seeds needed to apply Theorem 
\ref{teo:separation} while selecting nodes in regions $\Hc_l$ an $\Hc_r$,  thus the whole construction is consistent.

\section{Experimental validation}
Although our results hold asymptotically as $n \rightarrow \infty$,
we can expect to qualitatively observe the main effects predicted
by the analysis also in finite-size graphs. We will first
investigate the performance of graph matching 
algorithms in synthetic graphs generated according to our 
model of clustered networks, and then apply them to real 
social network graphs.  

\subsection{Synthetic graphs}
In this section we  consider bi-dimensional graphs having
$n = 10,000$, the sampling probability $s = 0.8$ and, unless otherwise specified, the average node degree in the ground-truth 
graph $D(n) = 30$. 

\begin{figure}[tbh]
\centering
\includegraphics[width=7cm]{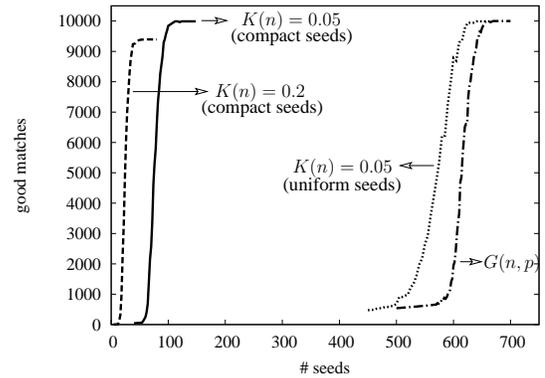}
\caption{Comparison of PGM performance (with $r=5$) 
in different networks with $n = 10,000$. 
Number of good matches (averaged over 1,000 runs) as a function of the number of seeds, chosen
either uniform or compact.}
\label{fig:1}
\end{figure}

Fig.~\ref{fig:1} reports the average number of correctly matched nodes
across $1,000$ runs of the PGM algorithm (using $r = 5$) in various cases, as function of the 
number of seeds. In each run, seeds are either chosen uniformly at random
among all nodes (label \lq uniform seeds'), or as a compact set around
one randomly chosen seed (label \lq compact seeds'). 
In our model of clustered graphs, we have fixed $\beta = 3$ (the decay
exponent of the edge probability beyond $C(n)$), and we consider
either $K(n) = 0.05$ or $K(n) = 0.2$. As reference, in the plot we also 
show  
the phase transition occurring (at about 600 seeds) when $\Tc$ is a $G(n,p)$ 
graph having the same average node degree. 
The plot confirms the wave-like nature of the identification process
as predicted by our analysis, namely: i) clustered networks (larger $K(n)$) 
can be matched starting from a much smaller seed set as compared to $G(n,p)$; ii) 
such huge reduction requires seeds to be selected within a small
sub-region of $\Hc$.

\begin{figure}[tbh]
\centering
\includegraphics[width=7cm]{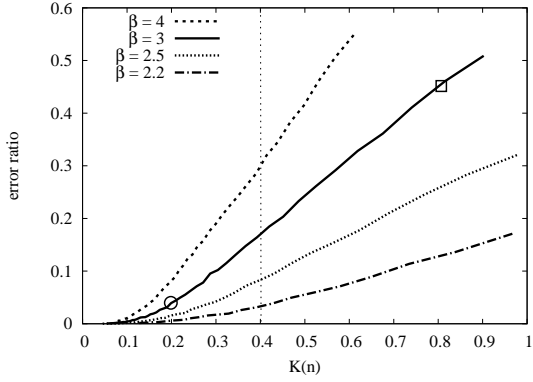}
\caption{Error ratio of PGM as a function of $K(n)$ for different values of 
$\beta$, starting from compact seeds.}
\label{fig:2}
\end{figure}

What the plot in Fig.~\ref{fig:1} does not clearly show (except 
for a rough estimate based on the maximum number of correctly matched nodes) 
is the error ratio incurred by the PGM algorithm, which is expected to become larger and larger
as we increase the level of clustering in the network. 
This phenomenon is confirmed by Fig.~\ref{fig:2}, which reports
the average error ratio (bad matches over all matches) incurred
by PGM as a function of $K(n)$, starting from a compact set of seeds.
In Fig.~\ref{fig:2} we have considered also different values of
$\beta$.
The little circle denotes the operating point
already considered for the left-most curve in Fig.~\ref{fig:1}, having an
error ratio of about 5\%. The plot reveals that the error ratio increases
dramatically when $K(n)$ tends to 1, confirming that PGM cannot be safely
applied in highly clustered networks. The effect of $\beta$ is more 
intriguing: smaller $\beta$'s produce fewer errors since 
generated network graphs tend to become more similar to $G(n,p)$, where
PGM is known to perform very well. As side-effect, smaller values of $\beta$
tend to slightly increase the percolation threshold (not shown in the plot).
For example, for $K(n) = 0.4$, the critical number of seeds  
(estimated from simulations) corresponding to $\beta = 2.2$,2.5,3,4 are 
equal to 11,15,24,45, respectively.

\begin{figure}[tbh]
\centering
\includegraphics[width=7cm]{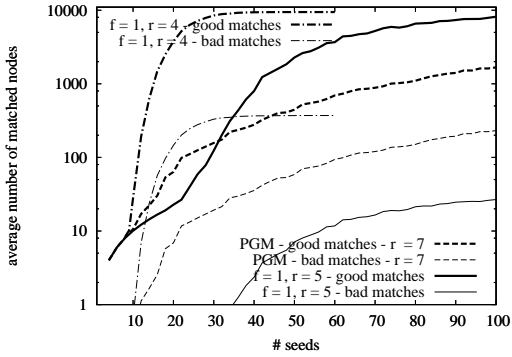}
\caption{Average number of good and bad pairs matched by different
algorithms for $K(n) = 0.8$, $\beta = 3$, starting from compact seeds.}
\label{fig:3}
\end{figure}

Next, we focus on the \lq hard' case corresponding to
the little square shown in Fig.~\ref{fig:2}, i.e., $K(n) = 0.8$,
$\beta = 3$. This case corresponds to networks having highly dense clusters,
where the performance of the original PGM algorithm 
is rather poor (error ratio about 50\%). 
Fig. ~\ref{fig:3}  shows the average number of nodes
matched by different algorithms as a function of the number of seeds: 
thick lines correspond to good matches, whereas thin lines (with the same line style)
refer to  bad matches produced by a given algorithm. 
For sake of simplicity, network de-anonymization is performed by
applying a simplified version of the algorithm proposed
and analysed in Section \ref{sec:dense}. This simple algorithm
consists in adopting PGM after having removed all graph edges shorter
than $x \cdot C(n)$.  
In the following, we will call this algorithm  \lq filtered PGM' and
we will label the corresponding curves in the plots  by \lq $f = <\!\!x\!\!>$'. 
We stress that  filtered PGM   provides qualitatively 
similar results to the performance of the algorithm in Section \ref{sec:dense}.

Looking at Fig. ~\ref{fig:3}, it is important  to remark  that in this scenario
the performance of the various algorithms is highly sensitive to the location of the  set of seeds
(in each run we uniformly select one seed among all nodes, 
and put all of the other seeds around it). Since we average the results over 1,000 runs, this explains
why all curves do not exhibit a 
sharp transition\footnote{We verified that, if we instead fix 
the very first seed across all runs, a sharp transition appears.  
However, the transition threshold changes as we vary the initial seed (results not shown here).}.
An average number of matched nodes equal to, say, 2,000,
must be given the following probabilistic interpretation: about
1/5 of (uniformly chosen) initial locations allow us to match almost all nodes (10,000),
while 4/5 of initial locations do not trigger the percolation effect.

Also, we note that the poor performance of standard PGM
cannot be fixed by just increasing the threshold $r$: using \mbox{$r=7$},
PGM still produces about $12\%$ error ratio, while requiring 
many more seeds (only about 2,000 nodes are matched on average 
starting from 100 seeds). Instead, filtered PGM, with $f = 1$ and $r=4$, requires very few seeds
to match almost all nodes, incurring about $3.7\%$ error ratio.
Using $f = 1$, $r = 5$, filtered PGM requires more seeds, but 
achieves as low as $0.3\%$ error ratio.

\begin{figure}[tbh]
\centering
\includegraphics[width=7cm]{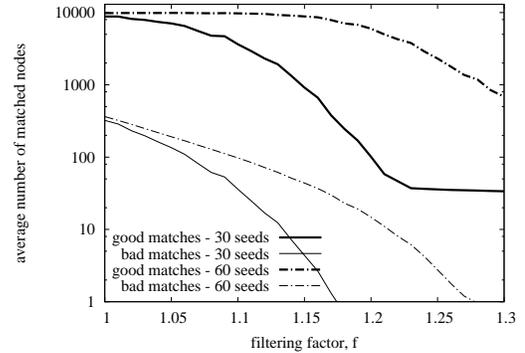}
\caption{Effect of varying the filtering factor $f$ for fixed $r = 4$ 
(scenario with $K(n) = 0.8$).}
\label{fig:4}
\end{figure}

Next, we fix $r$ and increase the filtering factor $f$
so as to diminish the number of errors while, however, reducing  the
average number of matched nodes (i.e., the probability to 
trigger percolation from a given seed set).
Fig. \ref{fig:4} illustrates this effect 
for $r=4$, in the case of two different seed set sizes, 
30 and 60. Having 60 seeds one could, for example, employ
$f = 1.1$ obtaining very high chance of percolation (almost 100\%)
and small error ratio (around $1\%$). 

\begin{table}
\label{tab:inv}
\begin{center}
\caption{Combinations of parameters achieving error ratio 3\%, percolation probability 50\%}
\vspace{1mm}
\begin{tabular}{|c|c|c|}
\hline
average node degree  & f & \# seeds  \\
\hline
36  & 1.1 & 22 \\
\hline
45  & 1.2 & 24 \\
\hline
53  & 1.3 & 28 \\
\hline 
64 & 1.4 & 32 \\
\hline
\end{tabular}
\end{center} 
\vspace{-8mm}
\end{table}

Alternately, we can fix a desired error ratio and average number of matched nodes
(i.e., the probability to trigger large-scale percolation), 
and look for the filtering factor and  number of seeds
that let us  achieve these goals. Table \ref{tab:inv} reports
an example of this numerical exploration, in which we vary 
the average degree of the nodes in $\Tc$ corresponding to each
examined scenario (the average degree can be increased, for fixed
$K(n) = 0.8$, by increasing $C(n)$). 
The results in Table \ref{tab:inv}  validate, at least
qualitatively, the counter-intuitive theoretical predictions in Table \ref{tab:res}: as we increase 
$C(n)$ (and thus the average node degree), the seed set size necessary 
to achieve a desired matching performance increases as well.

\subsection{Real social graphs} 
We consider a real graph derived from the Slovak social network 
Pokec \cite{takac}. The public data set, available at \cite{pokec}, is a directed graph
with 1,632,803 vertices, where nodes are users of Pokec and directed edges represent 
friendships. Since the original graph contains too many vertices 
for our computational power, and since we would like to isolate the impact of
clustering from the effect of long-tailed degree distributions, 
we considered only vertices having: i) in-degree larger than 20; 
ii) out-degree smaller than 200. We ended up with a reduced graph having
$n = 133,573$ nodes, average (in or out) degree 40.8 and clustering coefficient 0.1.
We use this graph as our ground-truth, and employ an edge sampling probability $s = 0.8$.
Notice that we maintain the direct nature of the edges, since 
all considered algorithms immediately apply
to direct networks as well \footnote{In direct networks, counters 
of matchable pairs are incremented only by using outgoing edges from 
matched pairs.}.

\begin{figure}[tbh]
\centering
\includegraphics[width=7cm]{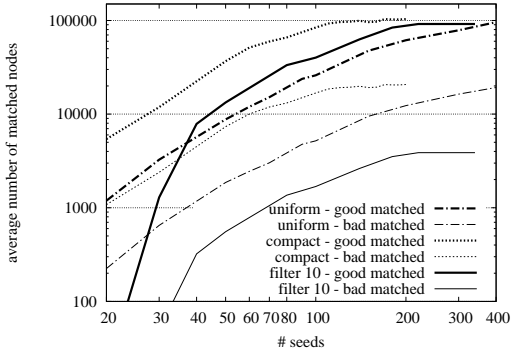}
\caption{Performance of matching algorithms in a subset of the 
friendship graph of the social network Pokec.}
\label{fig:pokec}
\end{figure}

Fig. \ref{fig:pokec} shows the performance of the different algorithms 
using threshold $r = 6$. As before, curves labelled \lq uniform' refer to the PGM
algorithm in which seeds are selected uniformly at random among the nodes.
Curves labelled \lq compact' refer to the PGM algorithm in which seeds 
are chosen among the closest neighbors of a uniformly selected node.  
Curves labelled \lq filter 10' differ from the previous one in that
the edges connecting each node to its nearest 10 neighbors are not used
by the algorithm. We emphasize that a $G(n,p)$ having the same number of nodes 
and average degree would require $a_c = 5,783$ seeds, according to (\ref{ac}).
In contrast, all considered algorithms require much fewer seeds to match
almost all nodes, confirming that real social networks are much simpler 
to de-anonymize than $G(n,p)$. 
In particular, the uniform variant requires about 300 seeds to match
on average more than 100,000 nodes, but incurs a quite large error ratio (about 17\%). 
The compact variant reduces this number roughly by a factor 3, but produces
the same error ratio. At last, the filtered variant requires slightly
more seeds than the compact one, but it allows to lower down the error ratio
to about 4\%.
The above results confirm the crucial performance improvement
that can be obtained by jointly: i) starting from a compact set of seeds 
(to exploit the wave-propagation effect), ii) carefully discarding edges connecting nodes 
to their local clusters (to limit the errors).

\section{Conclusions}
We focused on the effect of  node clustering on social graph  
de-anonymization. 
We defined a general model for network graphs that can   
represent  different 
levels of node clustering. Then we  
designed 
de-anonymization algorithms and
analysed their performance by using bootstrap percolation. 
Our theoretical results highlight that clustering significantly helps to reduce
the minimum seed set size required for network de-anonymization, and
that our algorithms can successufully limit the error rate of the
de-anonymization procedure.  
Our findings were confirmed by numerical experiments on synthetic and
real social graphs.


%


\begin{thebibliography}{99}
{\footnotesize


\bibitem{Narayanan} A. Narayanan, V. Shmatikov, ``De-anonymizing
social networks,'' {\em IEEE Symposium on Security and Privacy}, 
2009.

\bibitem{daniel}
P. Pedarsani, D.-R. Figueiredo, M. Grossglauser, ``A Bayesian method for matching two similar graphs 
without seeds,'' {\em IEEE Allerton} 2013.

\bibitem{peng} W. Peng, F. Li, X. Zou, J. Wu, ``A two-stage deanonymization attack against 
anonymized social networks,'' {\em IEEE Trans. on Computers}, 63(2), 2014.   

\bibitem{lattanzi} N. Korula, S. Lattanzi,  ``An efficient
  reconciliation algorithm for social networks,'' 
{\em PVLDB}, 
2014.

\bibitem{pedarsani}
P. Pedarsani, M. Grossglauser, ``On the privacy of anonymized networks,''
{\em SIGKDD}, 
2011.  

\bibitem{Grossglauser}
L. Yartseva, M. Grossglauser, ``On the performance of percolation
graph matching,''  {\em COSN},
2013.

\bibitem{Janson} S. Janson, T. Luczak, T. Turova, T. Vallier, ``Bootstrap percolation on the random
graph $G_{n,p}$,'' {\em The Annals of Applied Probability,} 22(5), 2012.

\bibitem{nostro-infocom} C.F. Chiasserini, M. Garetto, E.Leonardi,
``De-anonymizing scale-free social networks by percolation graph matching,''
{\em INFOCOM}, 2015.

\bibitem{tobias} K. Bringmann, T. Friedrich, A. Krohmer,
``De-anonymization of heterogeneous random graphs in quasilinear time,''
{\em  22nd Annual European Symposium on Algorithms, ESA'14.}









  





\bibitem{Penrose} M. Penrose, {\em Random Geometric Graphs,} Oxford
University Press, 2003.

\bibitem{health} Add health public data set, wave I (online) \url{http://www.cpc.unc.edu/projects/addhealth}

\bibitem{takac}
L. Takac, M. Zabovsky, ``Data analysis in public social networks," {\em Int. Scientific Conf. and Int. 
Workshop Present Day Trends of Innovations}, 2012. 

\bibitem{pokec} Pokec network dataset - KONECT, (website) \url{http://konect.uni-koblenz.de/networks/soc-pokec-relationships}


}
\end{thebibliography}

%
%

\appendix
\section{Concentration Inequalities from Penrose}
\label{penrose}
\begin{lemma}
Let $H(b)=1-b-b\log b$ for $b>0$.
Suppose $n \in \mathbb{N}$ $p \in (0,1)$ and $0\le k\le n$ let $\mu=n p$ if $k\le \mu$ then:
\begin{equation} \label{sotto}
 P(\text{Bin(n,p)}\le k)\le \exp \left(-\mu H\Big(\frac{k}{\mu}\Big)\right)
\end{equation}

if $k> \mu$ then:
\begin{equation} \label{sopra}
 P(\text{Bin(n,p)}\ge k)\le \exp \left(-\mu H\Big(\frac{k}{\mu}\Big)\right)
\end{equation}
if $k>e^2 \mu$ then 
\begin{equation} \label{soprasopra}
 P(\text{Bin(n,p)}\ge k)\le \exp \left(- \frac{k}{2} \log \frac{k}{\mu}\right)
\end{equation}

\end{lemma}

\section{Proof of Theorem \ref{teo:separation}}
\label{appA}

Without loss of generality, let us focus on $\Gc_1$ and 
let us consider a node $i\in  \Hc_{\text{in}}(\alpha,\delta)$. By construction,  
the number of seeds that are neighbors  of $i$ on $\Gc_1$ 
is given by  $S_i=\sum_{\sigma\in \Ac_0} X_{i\sigma}S^1_{i\sigma}\ge_{st} Y_i\ge_{st} Y$
where 
\[Y_i=\text{Bin}(a_0, sK(n) f( \max_{\sigma \in \Ac_0}||\xb_i
-\xb_{\sigma}||))\] 
and 
$Y=\text{Bin}(a_0, sK(n)(1+\delta)\alpha)$, with $ \mathbb{E}[Y]= s
K(n)(1+\delta)\alpha a_0$. 
Now, using the inequalities 
reported in Appendix~\ref{penrose},  
we can bound: 
\begin{multline}
P\left( Y_i<\alpha s K(n) a_0 \right)\le  \exp\left(- \mathbb{E}[Y_i] H\Big(\frac{\alpha s K(n) a_0}{\mathbb{E}[Y_i]}\Big)\right) \\
\le  \exp\left(- (1+\delta)\alpha s K(n) a_0 H\Big(\frac{1}{1+\delta}\Big)\right)
\end{multline}
with $H(b)=1-b +b\log b$. 
 
If we consider jointly all nodes in $\Hc_{\text{in}}(\alpha,\delta)$ and we denote with $N_{\text{in}}$
their number, 
we can bound the probability that every  node in  $\Hc_{\text{in}}(\alpha,\delta)$ is accepted  with:
\begin{multline} \label{probDin}
P\left( \text{all nodes in $\Hc_{\text{in}}$ are accepted} \mid  N_{\text{in}} \right) \\
\le 1- N_{\text{in}}\exp\left(- (1+\delta)\alpha sK(n)  a_0 H\Big(\frac{1}{1+\delta}\Big)\right),
\end{multline}
with  \eqref{probDin} that tends to 1 if 
$\log N_{\text{in}} - (1+\delta)\alpha
sH\Big(\frac{1}{1+\delta}\Big)K(n) a_0\to -\infty$. This can be enforced by opportunely  setting $a_0=\Omega\left(\frac{ \log N_{\text{in}}}{K(n)}\right)$.
Since by construction $|\Hc_{\text{in}}|>C^k(n)\ge\frac{log n}{n}$, we have w.h.p. $N_{\text{in}}   \le 2n|\Hc_{\text{in}}|$
 by standard concentration results (See Lemma \ref{lemmanumnodesinarea})
 ). As a consequence,  w.h.p. 
$$ P\left( \text{all vertices in $\Hc_{\text{in}}$ are accepted} \right) \to 1$$ provided that $a_0$ is opportunely chosen,  with:\\ 
$a_0=\Omega\left(\frac{\log(nC^k(n))}{K(n)} \right)$.


Then we focus on the nodes in $\Hc_{\text{out}}(\alpha, \delta)$ and we show that
all those nodes are jointly rejected.
 Conceptually we repeat the same approach as before, however, the argument is made  slightly more complex  by the fact that, to achieve tight  bounds on the probability that all nodes in $\Hc_{\text{out}}(\alpha,\delta)$ are jointly rejected, we need to   partition  $\Hc_{\text{out}}(\alpha,\delta)$ into smaller sub-regions containing nodes, which lie at  similar distance from the seeds.

Assuming $\delta< \frac{e^2-1}{e^2}$, we define $\Hc^1_{\text{out}}=\Hc^1(\alpha, \frac{e^2-1}{e^2})\subset \Hc_{\text{out}}(\alpha, \delta)$ and 
  $\Hc^0_{\text{out}}(\alpha,\delta) = \Hc_{\text{out}}(\alpha,\delta)
  \setminus \Hc^1_{\text{out}}$. Furthermore,  
we partition $\Hc^1_{\text{out}}$ into disjoint sub-regions, i.e.,  $\Hc^1_{\text{out}} = \cup_{h\ge 1}  \Hc^{1,h}_{\text{out}}$,
with $ \Hc^{1,h}_{\text{out}}=  \Hc_{\text{out}}(\frac{\alpha, h^\beta  e^2 -1}{h^\beta  e^2})\setminus 
\Hc_{\text{out}}(\alpha, \frac{(h+1)^\beta  e^2-1}{(h+1)^\beta  e^2})$.
Now,  given a vertex $i$    in $\Hc^{0}_{\text{out}}$   
 ($\Hc^{1,h}_{\text{out}}$),   the number of its neighbor seeds  $S_i$ on $\Gc_1$ can be bounded from above
 by a $\text{Bin}(a_0,sK(n)(1-\delta)\alpha)$  \Big($\text{Bin}(a_0, \frac{sK(n)}{h^\beta e^2}\alpha)$\Big).
Furthermore, by  elementary geometrical arguments, it can be shown that:
i) $|\Hc^{0}_{\text{out}}|=\Theta(C^k(n))$, 
ii)  $|\Hc^{1,1}_{\text{out}}|= \Theta(C^k(n))$   and 
iii)  $ \Hc_{\text{out}}^{1,h} = \Theta(  h^{k-1} \Hc_{\text{out}}^{1,1}$).

Denoted with $N^{0}_{\text{out}}$ and 
 $N^{1,h}_{\text{out}}$ the number of nodes in  $\Hc^{0}_{\text{out}}$ and  $\Hc^{1,h}_{\text{out}}$, respectively, by 
 exploiting again  the inequalities in 
 Appendix~\ref{penrose} w.h.p. we have:
\begin{multline}
P\left( \text{all nodes in $\Hc^0_{\text{out}}$ are rejected}\right) 
\le \\ 1- N^0_{\text{out}}\exp\left(- (1-\delta)\alpha sK(n) a_0
  H\Big({1-\delta}\Big)\right) \to 1  \,. \nonumber
\end{multline}
The above expression holds under the assumption that
$a_0=\Omega\left(\frac{\log(nC^k(n))}{K(n)} \right)$.  
Indeed, we remark that   $N^0_{\text{out}}\le 2n |\Hc^0_{\text{out}}|=\Theta(n C^k(n))$ w.h.p.
At last, 
\begin{multline}
P\left( \text{all nodes in $\Hc^1_{\text{out}}$ are  rejected}  \right) \\
\le 1-\sum_{h=1}^{\infty} N^{1,h}_{\text{out}}\exp\left(- \frac
  {\alpha s K(n) a_0}{2}(\beta \log h +2)  \right) \,. \nonumber
\end{multline}
For every $h$, $ N^{1,h}_{\text{out}}\le 2 n |\Hc^{1,h}|=\Theta( n
h^{k-1} C^k(n))$; also, the number of sub-regions of
$\Hc^1_{\text{out}}$ is $O(n/C^k(n))$. Thus,  w.h.p we have  that jointly on
all $h$'s, the number of nodes in these sub-regions can be bounded by  $ 2 n |\Hc^{1,h}|$. Under the  assumption  that 
$ a_0=\Omega\left(\frac{\log(nC^k(n))}{K(n)} \right)$, it can be easily shown  that $P\left( \text{all nodes in $\Hc^1_{\text{out}}$ are rejected}  \right)\to 1.$


\section{Proof of Theorem \ref{maintheo}}
\label{appB}

The following proof uses some notation  that has been introduced in \cite{Grossglauser} and that  here 
is omitted for brevity (the reader may also refer to Appendix \ref{app:theorem2} for a more detailed description of the PGM algorithm and associated notation). 

For any two vertices  $i\in \Mc_l$  
and $j\in \Mc_r$,    
let $X_{ij}$ be the Bernoulli random variable  that represents the presence of  an edge $(i,j)\in  \Ec$.
 By construction, $ Ber(p_{\min}) \le_{st}  X_{ij} \le _{st}
 Ber(p_{\max})$. I.e., two variables $\underline{X}_{ij}$ and
 $\overline{X}_{ij}$, with distribution, respectively,   $ Ber(p_{\min})$
and  $ Ber(p_{\max})$, can be defined on the same probability space as $X_{ij}$ such that 
 $\underline{X}_{ij} \le X_{ij} \le \overline{X}_{ij}$ point-wise.


We consider the corresponding pairs graph  $\Pc(\Tc)$, which is, 
by construction,  
composed of all the pairs of  vertices   residing in  $\Mc_l$ and
$\Mc_r$ and of the
 edges   connecting  pairs of vertices  in   $\Mc_l$  with pairs of vertices in  $\Mc_r$.
We denote by $\Pc_l$ and $\Pc_r$,  respectively, the set of  pairs
of $\Pc(\Tc)$, whose vertices lie in $\Mc_l$ and $\Mc_r$.  
Observe that, given two good pairs $[i_1,i_2] \in \Pc_l$ and
$[j_1,j_2] \in \Pc_r$,  the presence of an edge in $\Pc(\Tc)$ is
associated with  
the random variable:
\[
 Y_{[i_1,i_2], [j_1,j_2]}= X_{ij}X_{ij}S^1_{ij}S^2_{ij}= X_{ij}^2S^1_{ij}  S^2_{ij}
\]
where $S^1_{ij}$  and $S^2_{ij}$ are mutually independent $Ber(s)$ r.v's,  which are in turn independent of $X_{ij}$.
By construction,  $p_{\min}s^2 \le \mathbb{E}[Y_{[i_1,i_2], [j_1,j_2]}]\le p_{\max}s^2$.
Instead, given two bad pairs $[i_1,k_2]\in \Pc_l$  and $[j_1,l_2] \in \Pc_r$, 
$ Y_{[i_1,k_2], [j_1,l_2]}= X_{ij}X_{kl} S^1_{ij} S^2 _{kl}$,  
with $p^2_{\min}s^2 \le \mathbb{E}[Y_{[i_1,k_2], [j_1,l_2]}]\le p^2_{\max}s^2$.
Finally, if we consider   one good pair and  one bad pair (e.g., 
$[i_1,i_2]\in \Pc_l$  and  $[j_1,k_2]\in \Pc_r$), 
$ Y_{[i_1,i_2], [j_1,k_2]}= X_{ij}X_{ik} S^1_{ij} S^2 _{ik}$,  
with $p^2_{\min}s^2 \le \mathbb{E}[Y_{[i_1,i_2], [j_1,j_2]}]\le
p^2_{\max}s^2$. 

Recall that 
we assume that two seed sets,   
$\Ac^l_0 \in \Pc_l$ and  $\Ac^r_0 \in \Pc_r$ (with
$|\Ac^l_0|=|\Ac^r_0|$), are available.  
On  $\Pc(\Tc)$ we run  the PGM algorithm  \cite{Grossglauser}, opportunely modified, as follows.  
At every time step $t$, we extract uniformly at random one  pair
$\zv^l(t)=[z^l_1,z^l_2]_t \in  \Ac^l_{t-1}\setminus \Zc^l_{t-1}$ and
$\zv^r(t)=[z^r_1,z^r_2]_t\in \Ac^r_{t-1}\setminus \Zc^r_{t-1}$,  
 adding a mark to all the neighbor pairs in $\Pc_r$ and $\Pc_l$,
 respectively. In other words, matched pairs in $\Pc_l$ contribute to
 the mark of pairs in   $\Pc_r$ and vice versa.
Thus,  for a generic node  pair  $[i_1,j_2] \in 
\Pc_r\setminus \Zc^r_t$,  
marks are updated according to the  iteration:   $M^r_{[i_1,j_2]} (t)= M^r_{[i_1,j_1]}(t-1) +  Y_{\zv^l(t) ,[i_1,j_2]}$.
Similarly, for $[i_1,j_2] \in \Pc_l$ marks are updated according to    $M^l_{[i_1,j_2]} (t)= M^l_{[i_1,j_2]}(t-1)+  Y_{[i_1,j_2],\zv^r(t)}$.
For the rest, the algorithm proceeds exactly as described in Section \ref{sec:notmodel}.

Now, it is important to observe that marks of pairs on the RHS
 of the graph 
evolve exactly as the marks of a coupled PGM that operates  over  a
pairs graph $\Pc_R$ defined as follows. Denote the generic pair by
$[*_1,*_2]$; then $\Pc_R$ is a graph insisting on the set of  nodes 
$\Mc_r$ and in which 
the presence of edge $(\zv^r(t), [*_1,*_2])$,    
 for any $[*_1,*_2] \in \Pc_r\setminus \Zc^r_t$,   is dynamically unveiled at time $t$ by observing variable 
 $X_{z^l_1(t) *_1}X_{z^l_2(t) *_2} S^l_{z^l_1(t) *_1} S^r_{z^l_1(t)
   *_2}$. 
In other words, the edges originated from $\zv^l(t)$ are replaced by
the edges originated from $\zv^r(t)$ and viceversa. 

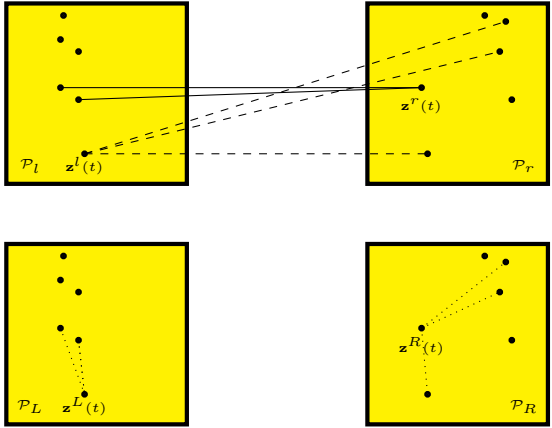
\begin{figure}[t.h]
\begin{center}
\begin{tikzpicture}[scale=0.8]
 \SetUpEdge[lw         = 1.5pt,
            color      = black,
            labelcolor = white]
  \GraphInit[vstyle=Normal] 

\draw [fill=  yellow,  ultra thick]  (0,4) rectangle (3,7);
\draw [fill= yellow,  ultra thick]  (6,4) rectangle (9,7);
\draw [fill=  yellow,  ultra thick]  (0,0) rectangle (3,3);
\draw [fill= yellow,  ultra thick]  (6,0) rectangle (9,3);

\draw [fill=black,ultra thick] (1.3,4.5) circle (0.02 cm);
\node at (1.3,4.3) { \tiny $\zv^l(t)$};

\draw [fill=black,ultra thick] (1.2,5.4) circle (0.02 cm);
\draw [fill=black,ultra thick] (0.9,5.6) circle (0.02 cm);
\draw [fill=black,ultra thick] (0.95,6.8) circle (0.02 cm);
\draw [fill=black,ultra thick] (1.2,6.2) circle (0.02 cm);
\draw [fill=black,ultra thick] (0.9,6.4) circle (0.02 cm);

\draw [fill=black,ultra thick] (7,4.5) circle (0.02 cm);
\draw [fill=black,ultra thick] (8.4,5.4) circle (0.02 cm);

\draw [fill=black,ultra thick] (6.9,5.6) circle (0.02 cm);
\node at (6.9,5.3) { \tiny $\zv^r(t)$};

\draw [fill=black,ultra thick] (7.95,6.8) circle (0.02 cm);
\draw [fill=black,ultra thick] (8.2,6.2) circle (0.02 cm);
\draw [fill=black,ultra thick] (8.3,6.7) circle (0.02 cm);

\draw[dashed,thin]  (1.3,4.5) -- (7,4.5);
\draw[dashed,thin]  (1.3,4.5) -- (8.2,6.2);
\draw[dashed,thin]  (1.3,4.5)  -- (8.3,6.7);

\draw[solid,thin]  (1.2,5.4) -- (6.9,5.6) ;
\draw[solid,thin]  (0.9,5.6) --  (6.9,5.6);

\draw[dotted,thin]  (1.2,1.4) -- (1.3,0.5); 


\draw [fill=black,ultra thick] (1.3,0.5) circle (0.02 cm);
\node at (1.3,0.3) { \tiny $\zv^L(t)$};

\draw [fill=black,ultra thick] (1.2,1.4) circle (0.02 cm);
\draw [fill=black,ultra thick] (0.9,1.6) circle (0.02 cm);
\draw [fill=black,ultra thick] (0.95,2.8) circle (0.02 cm);
\draw [fill=black,ultra thick] (1.2,2.2) circle (0.02 cm);
\draw [fill=black,ultra thick] (0.9,2.4) circle (0.02 cm);

\draw [fill=black,ultra thick] (7,0.5) circle (0.02 cm);
\draw [fill=black,ultra thick] (8.4,1.4) circle (0.02 cm);
\draw [fill=black,ultra thick] (6.9,1.6) circle (0.02 cm);
\node at (6.9,1.3) { \tiny $\zv^R(t)$};

\draw [fill=black,ultra thick] (7.95,2.8) circle (0.02 cm);
\draw [fill=black,ultra thick] (8.2,2.2) circle (0.02 cm);
\draw [fill=black,ultra thick] (8.3,2.7) circle (0.02 cm);

\draw[dotted,thin]  (6.9,1.6) -- (7,0.5);
\draw[dotted,thin]  (6.9,1.6) -- (8.2,2.2);
\draw[dotted,thin]  (6.9,1.6)  -- (8.3,2.7);

\draw[dotted,thin]  (1.2,1.4) -- (1.3,0.5) ;
\draw[dotted,thin]  (0.9,1.6) --  (1.3,0.5);

\node at (0.4,0.3) { \tiny $\Pc_{L}$};
\node at (8.6,0.3) { \tiny $\Pc_{R}$};

\node at (0.4,4.3) { \tiny $\Pc_l$};
\node at (8.6,4.3) { \tiny $\Pc_r$};
\end{tikzpicture}
\end{center}
\caption{Graphical representation  of the PGM evolution over coupled graphs.\label{fig:overview-proof}}
\end{figure}

Furthermore, we make the following observations.

(i) We assume that the sequence of matched pairs $\{\zv^R_t\}_t \in \Pc^{(R)}$ exactly corresponds to the sequence of 
matched pairs $\{\zv^r(t)\}_t  \in \Pc_r$, i.e., $\zv^r(t)=\zv^R(t)$ at every $t$.
This is made possible by the fact that given $\Zc^r_{t-1}=\Zc^R_{t-1}$,
marks collected by  every unmatched pair  
in the two graphs at time $t$ exactly correspond. 

(ii) Our construction is consistent since edges between pairs are unveiled
 only once,  specifically at the time  at which 
the first between the two edge endpoints in $\Pc_R$
is placed in $\Zc^R_t=\Zc^r_t$. Since then, the edge is replaced
with an edge between two pairs that are both in  $\Pc_R$, hence it
will not be used again.

(iii) $\Pc_R$  is isomorphic to a pairs graph originated  
by a  generalized Erd\"{o}s--R\'{e}nyi  graph  $\Tc^R$,  in which
the presence of 
every  edge $(\zv^r(t), *)$ can be represented by a Bernoulli r.v. and
the probability that the edge is added to  the graph takes values in the range
  $[p_{\min}, p_{\max}]$ and is 
independent of other edges. 
  Indeed, observe that the presence of an edge in $\Pc_R$ deterministicaly 
  corresponds to the presence of the corresponding edge in
  $\Pc(\Tc)$.  
  Furthermore, by costruction, different edges in $\Pc_R$  correspond 
  to different edges in $\Pc(\Tc)$. 

The same observations hold when we consider the evolution of the marks
of the pairs on the left hand side and a pairs graph $\Pc_L$, which  is originated from a coupled generalized  Erd\"{o}s--R\'{e}nyi 
graph  $\Tc^L$ with
same properties as  $\Tc^R$. 

Now, clearly  $G(m,p_{\min}) \le_{st}  \Tc^R \le_{st}  G(m,p_{\max})$
and   $G(m,p_{\min}) \le_{st}  \Tc^L \le_{st}  G(m,p_{\max})$, i.e.,
$\Tc^R$  ($\Tc^L$) 
can be obtained by opportunely thinning  a graph $G(m,p_{\max})$, while a graph $G(m,p_{\min})$ can be obtained by 
opportunely thinning $\Tc^R$  ($\Tc^L$).
 Then we  invoke Theorem \ref{propLU} 
to conclude our proof and show that our algorithm correctly percolates
over $\Tc^R$ and $\Tc^L$ and, thus,  over the original 
bipartite $\Tc$.


\section{Extracting nodes from a defined region $H_0$} \label{app:extactingnodes}


Our matching procedure 
requires  to extract (select) nodes 
that lye in a defined  region $\Hc_0$.

Clearly, to   extract nodes  lying in a defined  region without
errors, 
it is necessary to have direct access  to   vertices' positions.
However,  our algorithm has access only to graphs  $\Gc_1$ and $\Gc_2$ (i.e., their adjacency matrix), and thus 
it extracts nodes based on  ``estimated'' positions/distances (i.e. according to Theorem~\ref{teoseparation} 
or Proposition~\ref{propdist2}.

Thus, if we  extract   nodes on the basis of their estimated position, 
we will necessarily incurr in  some error: some nodes in $\Hc_0$  will
not be  selected while 
others lying outside  $\Hc_0$  will be selected.
We denote with $\Pc(\Hc_0)$  the set of pairs whose nodes lye in $\Hc_0$ and with $\hat{\Pc}(\Hc_0)$ the set of pairs 
composed by nodes that are extracted.

We need to devise a smart strategy that extracts nodes  while guaranteeing that  the following
three conditions are satisfied:
\begin{enumerate}
\item Only good pairs  formed by  vertices  whose actual location is in $\Hc_0$   (i.e. good pairs in $\Pc(\Hc_0)$) are extracted; 
\item A finite fraction (bounded away from 0) of good  pairs of   $\Pc(\Hc_0)$  is estracted (i.e., included in $\hat{\Pc}(\Hc_0)$; 
\item  The following situation occurs with negligible probability:
 a bad pair $[i_1,j_2]$  is included in $\hat{\Pc}(\Hc_0)$ while none of the pairs
 $[i_1,i_2]$ and $[j_1,j_2]$ are included.
\end{enumerate}
The third condition ensures that every selected bad pair is in
 conflict with at least one good pair in the set,
thus it will not  be matched by the PGM algorithm 
when it (eventually) reaches the threshold. 
Below, we show how conditions 1) 2) and 3) can be easily
guaranteed. For simplicity, we restrict our attention 
to spheric regions, although the 
same argument can be applied to regions of any shape.

We first  introduce this preliminary result.

\begin{proposizione} \label{prop:imperfect}
Assume that position of nodes (lenght of edges) are estimated with a bounded error $\Delta$.
Then, given  a spheric  region $\Hc_0$ whose side is not smaller than $7 \Delta$, it is possible to extract a 
set of nodes from $\Gc_1$ and $\Gc_2$ (and consequently to define $\hat{\Pc}(\Hc_0)$) satisfying conditions 1), 2) and 3).
\end{proposizione} 
\IEEEproof 

We select nodes as
follows. We  partition  region $\Hc_0$ into three disjoint sub-regions. An inner spheric region  of radius $3 \Delta$ 
co-centered within $\Hc_0$, an intermediate annulus-shaped region 
with external radius equal to  $5 \Delta$, and  a remaining outer  region.

The idea is to extract only those pairs of vertices whose estimated position falls in 
either the inner or  the intermediate region, under the additional condition 
that only pairs for which at least one vertex falls in the inner
region  are extracted.  
This expedient  implies that $[i_1,j_2]$ is 
selected  only if   the estimated location of
$i_1$  ($j_2$) falls  in the inner region and the estimated position 
$i_2$  ($j_1$) falls in either the inner or the  intermediate region.
Clearly, the true position of $i_1$ ($j_2$) must necessarily lie in $\Hc_0$. Furthermore, all nodes whose true position falls 
in a 
spheric region of  radius $\Delta$ co-centered with $\Hc_0$  will be necessarily selected, thus conditions 1) and 2) are met w.h.p. 
as immediate consequence of  Lemma \ref{lemmanumnodesinarea}. 
Finally, 3) is necessarily met as result of the following argument. 
(i) Observe  that, for every  node $i$, the distance between
the estimated positions of $i_1$ and $i_2$ is by construction  smaller
than $2c_0C(n)$.  (ii) Then let us consider a selected bad pair $[i_1, j_2]$; without lack of generality, we can assume the estimated position of $i_1$ to lye in the inner region.
From consideration (i), the estimated position of $i_2$ must necessarily lye either in the inner or the intermediate region. 
(iii) As a result, the pair  $[i_1,i_2]$  is necessarily selected too by our algorithm.

\begin{proposizione}
The same approach can be  pursuit in the case of the application of Theorem \ref{teo:separation}  
to define the initial  set of vertices pairs $\Pc(\Nc)$  
so as to satisfy condition 3) (along with 1) and 2). 
\end{proposizione}

Indeed, 
in such a case the role of the inner region is played by $\Dc_{\text{in}}(\alpha_1 \delta)$, the role of intermediate region is played by       
 $\Dc_{\text{in}}(\alpha_2 \delta) \setminus \Dc_{\text{in}}(\alpha_1 \delta) $ while the role of outer region is played by 
$\Dc_{\text{out}}(\alpha_1 \delta) \setminus \Dc_{\text{in}}(\alpha_1 \delta)$.  
Indeed, by construction, if a vertex  $i_1$ is accepted by adopting a threshold $\alpha_1$, the corresponding vertex  
  $i_2$  will be necessarily  accepted by adopting a threshold $\alpha_2$.  
\endIEEEproof


\section{Lemma \ref{lemmanumnodesinarea}  and Proof of Theorem \ref{teo-prop}}
\label{app:teo-prop}
\begin{lemma}  \label{lemmanumnodesinarea} 
The number $N_{\Hc_0}$ of nodes falling in  a region $\Hc_0$  satisfies  
$ \frac{n}{2} |\Hc_0| <  N_{\Hc_0} < 2n |\Hc_0|$ w.h.p., as long  as $|\Hc_0|= \omega(\frac{1}{n})$.
In particular, if   $|\Hc_0|\ge c \frac{\log n }{n}$,   then $ \frac{n}{2} |\Hc_0| <  N_{\Hc_0} < 2n |\Hc_0|$ with a probability 
$1- O(n^{cH(1/2)})$.
\end{lemma}
\IEEEproof 
The proof immediately descends by applying  \eqref{sotto} and
\eqref{sopra} to  $N_{\Hc_0}= Bin( n,|\Hc_0|)$ with
$\mu=\mathbb{E}[N_{\Hc_0}]=n |\Hc_0|$. 
\endIEEEproof

\begin{figure}[t.h]
\begin{center}
\begin{tikzpicture}[scale=0.8]
 \SetUpEdge[lw         = 1.5pt,
            color      = black,
            labelcolor = white]
  \GraphInit[vstyle=Normal] 
 
\draw [fill=yellow,ultra thick] (0,0) circle (1 cm);
\draw [thin, dashed] (0,0) -> (-1,0);
\node at (-0.6,-0.2) {\tiny $\rho$};
\node at (0,2.1) { \tiny $\xb_i$};
\draw [fill=black,ultra thick] (0,2) circle (0.02 cm);
\draw [fill=black,ultra thick] (1.2,0) circle (0.02 cm);
\draw [thin] (0,2) circle (0.4 cm);
\draw [thin] (0,2) circle (0.8 cm);
\node at (-0.4, 3.1) {\tiny $h_1=3$};
\draw [thin] (0,2) circle (1.6 cm);
\node at (-0.4, 4.0) {\tiny $h_1=4$};
\draw [thin] (0,2) circle (3.2 cm);

\node at (1.2,0.1) {\tiny $\xb_j$};
\draw [thin] (1.2,0) circle (0.4 cm);
\draw [thin] (1.2,0) circle (0.8 cm);
\draw [thin] (1.2,0) circle (1.6 cm);
\node at (1.9, -1) {\tiny $h_2=3$};
\draw [thin] (1.2,0) circle (3.2 cm);
\node at (1.9, -2) {\tiny $h_2=4$};
\end{tikzpicture}
\end{center}
\caption{Graphical representation  of $\Dc_1(h_1)$  and  $\Dc_2(h_2)$\label{fig:fig1}}
\end{figure}
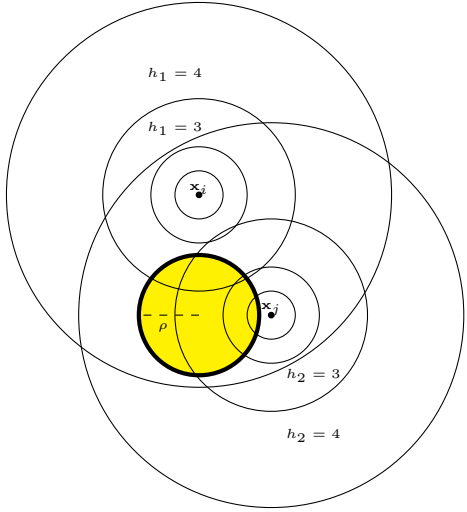


{\bf Proof of Theorem  \ref{teo-prop}.}
The proof of this proposition is based on the application of standard
concentration results, namely, Chernoff bound and inequalities reported in Appendix~\ref{penrose}.
For the sake of clarity, we restrict ourselves to consider the case $k=2$; the extension to a generic  $k$ is easy to obtain.

Consider a correct pair $[i_1,i_2]\in \Dc(0, \rho_1) \setminus \Dc(0,
\rho)$  whose location in $\Hc$ is denoted by $\xb_i$. 
We compute its  number of edges  with pairs in $\Dc(0, \rho)$,
 $N_i=  \sum_{l\in \Dc(0, \rho)} Y[i_1,i_2][l_1 l_2]] \ge  \sum_{l\in \Dc(0, \rho)\cap \Dc(\xb_i, C(n))  } Y[i_1,i_2][l_1 l_2]= 
\text{Bin}(N_{ \Dc(0, \rho)\cap \Dc(\xb_i, C(n)) }, K(n)) $  where $N_{ \Dc(0, \rho)\cap \Dc(\xb_i, C(n)) }$ denotes the number of nodes in 
$\Dc(0, \rho)\cap \Dc(\xb_i, C(n))$.  

As an  immediate consequence of Lemma \ref{lemmanumnodesinarea}, $N_{ \Dc(0, \rho)\cap \Dc(\xb_i, C(n)) } > \frac{n}{2}  | \Dc(0, \rho)\cap \Dc(\xb_i, C(n))| $ with a probability $1- O(n^{-2})$.  Then,
conditionally to  this relation,  
we have 
$\text{Bin}(N_{ \Dc(0, \rho)\cap \Dc(\xb_i, C(n)) }, K(n))>  \frac{n}{2}  | \Dc(0, \rho)\cap \Dc(\xb_i, C(n))|\frac{K(n)}{2}$ with a probability $1- O(n^{-\gamma})$ with $\gamma>0$, as  it can be immediately shown  by applying $\eqref{sotto}$.  

As a consequence,  our algorithm successfully identifies almost all
good pairs in $\Dc(0, \rho_1) \setminus \Dc(0, \rho)$ 
(i.e.,  $N_{ \Dc(0, \rho)\cap \Dc(\xb_i, C(n)) }- o(N_{ \Dc(0, \rho)\cap \Dc(\xb_i, C(n)) })$)
with a probability $1- O(n^{-1})$, again, as a consequence of $\eqref{sotto}$ when applied to the number of matched nodes in 
$ \Dc(0, \rho)$.

Next, consider a  bad pair $[i_1,j_2]$ whose nodes $i$ and $j$ are located respectively in $\xb_i$ and $\xb_j$, with $|\xb_i|= \rho_i$  and   $|\xb_j|= \rho_j$.  
Let  $\Dc_1(h_1)= \Dc(\xb_i, 2^{h_1+1}C(n)) \setminus   \Dc(\xb_i, 2^{h_1}C(n)$ for $h_1\ge 1$ 
with $\Dc_1(0)=\Dc(\xb_i,2C(n))$  and 
$\Dc_2(h_2)= \Dc(\xb_j, 2^{h_2+1}C(n)) \setminus   \Dc(\xb_i,
2^{h_2}C(n))$ with  $\Dc_2(0)=\Dc(\xb_j,2C(n))$ (see Figure \ref{fig:fig1}).

Let 
$\Cc(h_1,h_2)=\Dc(0, \rho)\cap\Dc_1(h_1) \cap \Dc_2(h_2)$   for $h_1\ge 1$ and $h_2\ge 0 $. We have:
  $N_{[i_1,j_2]}=  \sum_{l\in \Dc(0, \rho)} Y[i_1,j_2][l_1 l_2]]  =
  \sum_{h_1} \sum_{h_2} \sum_{l\in \Cc(h_1,h_2) } Y[i_1,j_2][l_1 l_2]\le 
\sum_{h_1} \sum_{h_2} \text{Bin}(N_{  \Cc(h_1,h_2)}  , K^2(n)
2^{-\beta(h_1+h_2)} )$.  

Now, 
$\Cc(h_1,h_2)$  is a subset of both $\Cc({h_1})= \Dc(0, \rho) \cap \Dc_1(h_1)$ and 
$\Cc({h_2})=\Dc(0, \rho)\cap \Dc_2(h_2)$.  
Thus $N_{  \Cc(h_1,h_2)}\le \min(N_{  \Cc(h_1)}, N_{ \Cc(h_2)})$.
In addition,  by construction: 
$\Dc(0, \rho)\cap \Dc_1(h_1)=\emptyset$  if 
$h_1<h_1^{\min} \lceil \log_2 (1+ \frac{\rho_i- \rho}{C(n)}) \rceil $, or $h_1^{\max}>\lceil \log_2 (1+ \frac{\rho_i+ \rho}{C(n)}) \rceil$. Similarly,
$\Dc(0, \rho)\cap \Dc_2(h_2)=\emptyset$  if 
$h_2< h_2^{\min}\lceil \log_2 (1+ \frac{\rho_i- \rho}{C(n)}) \rceil $, or $h_2> h_2^{\max}=\lceil \log_2 (1+ \frac{\rho_i+ \rho}{C(n)}) \rceil$.

Hence,  
\begin{eqnarray}
N_{[i_1,j_2]}&\le&  \sum_{{h_1^{\min}}}^{h_1^{\max}}
\sum_{{h_2^{\min}}}^{h_2^{\max}} \text{Bin}( \min(N_{  \Cc(h_1)}, N_{
  \Cc(h_2)}) , \nonumber \\
&& K^2(n) 2^{-\beta(h_1+h_2)} )
\end{eqnarray}
Now $C(h_1)$ is by construction  a subset of $\Dc(0, \rho)$  as well as of $\Dc_1(h_1)$,  thus 
$N_{C(h_1)} \le \min(N_{\Dc(0, \rho)}, N_{\Dc_1(h_1)})$  
and similarly $N_{C(h_2)} \le \min(N_{\Dc(0, \rho)}, N_{\Dc_2(h_2)})$, thus:
\begin{multline}
N_{[i_1,j_2]}\le  \sum_{{h_1^{\min}}}^{h_1^{\max}} \sum_{{h_2^{\min}}}^{h_2^{\max}} 
\text{Bin}( \min(N_{\Dc(0, \rho)},  N_{ \Dc_1(h_1)
},N_{\Dc_2(h_2)}),\\   K^2(n) 2^{-\beta(h_1+h_2)}) \,.
\end{multline}

Note that $|\Dc(0, \rho)|=\pi \rho^2$  while  
$|\Dc_1(h_1)|\le 
\pi 2^{2(h_1+1)} C^2(n)$,  
and, similarly,  $|\Dc_2(h_2)|\le  \pi 2^{2(h_2+1)} C^2(n)$.
As a consequence, since all these defined regions are larger than $C^2(n)$,  
 from Lemma \ref{lemmanumnodesinarea}  we have that, uniformly on  $h_1$ and $h_2$,  the number of nodes in these 
 regions is not larger than $2n$ times the volume of the regions themselves. I.e.,
$N_{\Dc(0, \rho)}< 2n \pi \rho^2$, $N_{ \Dc_1(h_1)}\ge  2n\pi 2^{2(h_1+1)} C^2(n)$  
and $N_{\Dc_2(h_2) }< 
2n\pi (2^{2((h_2+1))}  C^2(n)$   with a probability $1- O(n^2)$.
Thus, by construction:
\begin{multline}
 \mathbb{E}[N_{[i_1,j_2]}]\le  \sum_{{h_1^{\min}}}^{h_1^{\max}} \sum_{{h_2^{\min}}}^{h_2^{\max}} 
\mathbb{E}[\text{Bin}( 2n  C^2(n) \\
\min\left(2^{2(h_1+1)} ,2^{2(h_2+1)}, \pi\frac{\rho^2}{C^2(n)} ) \right), K^2(n) 2^{-\beta(h_1+h_2)})] (1- O(n^{-2}))\\
+ n O(n^{-2})
\end{multline}
 Furthermore, $\min(2^{2(h_1+1)} ),2^{2(h_2+1)} )\le
 2^{2(h_1+h_2)+3}$. 
Then we can rewrite the previous expression as:
\begin{multline}
\mathbb{E}[N_{[i_1,j_2]}]\le  
\sum_{{h_1^{\min}}}^{h_1^{\max}} \sum_{{h_2^{\min}}}^{h_2^{\max}} \\
\mathbb{E}[Bin( 2n C^2(n) \min(2^{2(h_1+h_2)+3},\\ \pi \frac{\rho^2}{C^2(n)} ),  K^2(n) 2^{-\beta(h_1+h_2)} )]+ O(n^{-1}).
\end{multline}

Now, if $2^{2(h^{\min}_1+h^{\min}_2)+3}< \pi \frac{\rho^2}{C^2(n)}$,  we can bound:
\begin{multline}
\mathbb{E}[N_{[i_1,j_2]}]\le \\ \sum_{{h_1^{\min}}}^{h_1^{\max}} \sum_{{h_2^{\min}}}^{h_2^{\max}} 
\mathbb{E}[Bin( n  2^{2(h_1+h_2)+4},  K^2(n) 2^{-\beta(h_1+h_2)} )]+ O(n^{-1}) =\\
 n C^2(n) \sum_{{h_1^{\min}}}^{h_1^{\max}} \sum_{{h_2^{\min}}}^{h_2^{\max}} 
 2^{4+ h_1(2-\beta)+ h_2(2-\beta) }K^2(n)+ O(n^{-1})\le\\
 n C^2(n)2^{4+ (h_1^{\min}+h_2^{\max}(2-\beta) }K^2(n) \sum_{0}^{\infty} \sum_{0}^{\infty} 2^{h_1+h_2(2-\beta)}\\ +
O(n^{-1})=\\  
2n C^2(n)2^{3+ (h_1^{\min}+h_2^{\min})(2-\beta) }K^2(n)\left(\frac{1}{1-
  2^{2-\beta}}\right)^2+O(n^{-1}) \,.
\end{multline}
If, instead,  $2^{2(h^{\min}_1+h^{\min}_2)+3}> \pi \frac{\rho^2}{C^2(n)}$, with similar arguments 
we can bound:
\begin{multline}
\mathbb{E}[N_{[i_1,j_2]}]\le  2n\pi \rho^2  K^2(n) 2^{-(h_1^{\min}+h_2^{\min})\beta}(\frac{1}{1- 2^{-\beta}})^2+O(n^{-1}).
\end{multline}
Observe that,  in general, 
\[
\mathbb{E}[N_{[i_1,j_2]}]=O(n [C(n)]^2K^2(n))
\]
with   $\mathbb{E}[N_{[i_1,j_2]}]=\Theta(n C^2(n)K^2(n))$ only when $h_1^{\min} + h_2^{\min}$ is bounded.
As a consequence,  the bad pair  $[i_1,j_2]$ will not  reach
threshold $r=\Theta(nC^2(n)K^2(n))$  w.h.p, as  it can be immediately verified by applying 
Markov inequality. However, we need to show that jointly 
all bad pairs will   remain below the threshold with a probability 
$1- O(n^{-1})$. We  can prove this stronger property 
first by  deriving a   tighter bound for the  probability that a specific pair 
reaches the threshold, and then by  applying the union bound on all  pairs.

Considering again the bad  pair $[i_1, j_2]$,  
 $N_{[i_1,j_2]}$ can be rewritten as $N_{[i_1,j_2]}=\sum_{l \in
   \Hc,l\neq i,j} \ind_{l\in \Dc(0,\rho)} Y[i_1,i_2][l_1 l_2]] $,
 i.e., as a sum of  independent Bernoulli random variables. Thus, we can apply Chernoff inequality to bound its tail. 
Recalling that by construction $r\gg \mathbb{E}[N_{[i_1,j_2]}]$, we have:
\begin{multline}
P(N_{[i_1,j_2]}\ge r )\le e^r \left(\frac{ \mathbb{E}[N_{[i_1,j_2]}] }{r}\right)^r=\\
e^{r\left(1- \log \frac{r}{ \mathbb{E}[N_{[i_1,j_2]} ] }\right)}.
\end{multline}

From  the definition  of $r$,  it follows that   $r=c n C^2(n) K(n)$ with $c=\frac{|\Dc(0,\rho)|\cap 
|\Dc(\rho_1,C(n) )|}{4C^2(n)}>0$.
Thus, 
\begin{multline}
\log P(N_{[i_1,j_2]}\ge r ) \le cn C^2(n) K(n)\\
\left(1- \log \frac{1}{K(n)}- (\beta-2)(h_1^{\min}+ h_2^{\min})\log 2 +C_1 \right)
\end{multline}
where $C_1$ is   an opportune constant.
By assumption, $n C^2(n) K(n)\ge \log n$ and $K(n)=o((\log
n)^{-\gamma})$, hence  
\[
 \log P(N_{[i_1,j_2]}\ge r ) \le - c\log n \cdot \omega(1).
\]
Since for large $n$ we have $c\log n \cdot \omega(1)> 3 \log n$,
it turns out that every bad pair $[i_1,j_2]$, regardless the position of its vertices, 
reaches  threshold $r$ with a probability $O(n^{-3})$. 
By applying the union bound, we can claim that jointly all of such
pairs will remain below the threshold $r$ with a probability 
$O(n^{-1})$. 

\section{Proof of Theorem \ref{propLU}}
\label{app:theorem2}

\begin{algorithm}
\begin{algorithmic}[1]
\State $\Ac_0=\Bc_0=\Ac_0(n)$, $\Zc_0=\emptyset$
\While{$\Ac_t\setminus \Zc_t\neq\emptyset$}  
\State $t=t+1$
\State Randomly select a pair $[*_1,*_2]\in \Ac_{t-1}\setminus \Zc_{t-1}$  
and add one  mark to all neighbor pairs of  $[*_1,*_2]$ in
$\Pc(\Tc)$.  
\State Let  $\Delta \Bc_t$ be the set of  all  neighbor pairs  of   $[*_1,*_2]$ in $\Pc(\Tc)$ whose mark counter has reached threshold 
 $r$   at time $t$.
\State Construct   set $\Delta   \Ac_t \subseteq \Delta \Bc_t $ as follows. 
Order the pairs in $\Delta \Bc_t $ in an arbitrary way,
  select them sequentially and test them for inclusion in $\Delta A_t$
\If{the selected  pair in  $\Delta \Bc_t$  has no conflicting pair in $\Ac_{t-1}$ or $\Delta \Ac_t$}
\State Insert the pair in $\Delta \Ac_t$
\Else \State Discard it
\EndIf
\State $\Zc_t=\Zc_{t-1}\cup[*_1,*_2]$, $\Bc_t=\Bc_ {t-1}\cup\Delta
\Bc_t $,  $\Ac_t=\Ac_{t-1}\cup\Delta \Ac_t$
\EndWhile 
\State \Return $T=t$, $\Zc_T=\Ac_T$
\end{algorithmic}
\caption{\label{alg:PGM}The PGM algorithm}
\end{algorithm}

The proof we propose complements the one provided   in \cite{nostro-infocom}, which holds only   under the assumption  
$p_{\min}  \gg \sqrt{\frac{n^{-3/r-1}}{ s^2}}$.
Here, we restrict to the case  $p_{\min} = O\left(\sqrt{\frac{n^{-3/r-1}}{ s^2}}\right)$.
 With reference to PGM algorithm reported in Figure~\ref{alg:PGM}, we define: \vspace{-1mm}
\begin{itemize}
\item 
$\Bc_t(\Tc)$ as the set of  pairs in $\Pc(\Tc)$
 that at time step $t$ have already collected a least
$r$ marks. 
It is composed of good pairs $\GoodB_{t}(\Tc)$ and bad pairs $\BadB_{t}(\Tc)$; \vspace{-1mm}
\item 
$\Ac_t(\Tc)$ as the set of matchable pairs at time
   $t$. Similarly to $\Bc_t(\Tc)$, it comprises
   good pairs $\Good_t(\Tc)$ and bad pairs $\Bad_t(\Tc)$.
In general, $\Ac_t(\Tc)$ and $\Bc_t(\Tc)$ do not coincide as
  $\Bc_t(\Tc)$ may include conflicting pairs that are not present in
  $\Ac_t(\Tc)$; \vspace{-1mm}
\item 
$\Zc_t(\Tc)$ as the set of pairs 
that  have been matched up to time $t$. By construction, $|\Zc_t|=t$, $\forall t$.
\end{itemize}\vspace{-1mm}

Next, we define $T_{G_{p_{\min}}}=\min\{ t \,\mbox{\em s.t.} \, |\Ac_t (G(n_, p_{\min})|=t\}$ and 
  $T_{G_{p_{\max}}}= \min\{ t \,\mbox{\em s.t.} \, |\Ac_t (G(n_, p_{\max})|=t\}$. 
By Theorem~\ref{theorem:gross}, we have that both   
$T_{G_{p_{\min}}}$ and $T_{G_{p_{\max}}}$ are equal to $n-o(n)$. 
Then inductively on $t$,  $\forall t < \min(T_{G_{p_{\min}}}, T_{G_{p_{\max}}})$, w.h.p.:
\begin{equation} \label{neweq1}
| \BadB_t (\Gc_T) | \le | \BadB_t ((G(n, p_{\max})) | =\emptyset
\end{equation}
In (\ref{neweq1}), the inequality descends  by  monotonicity of sets  $\BadB_t $  with respect  to ``$\le_{\text{st}}$''. The following equality 
descends from Corollary 1 in \cite{nostro-infocom} applied to
$G_{p_{\max}}$. We remark that, under our assumption on $p_{\min}$ and  $p_{\max}$,  we have $t_0=T$ in Corollary 1 in \cite{nostro-infocom},  along with:
\begin{align} \label{neweq}
|\Ac_t (\Gc_T)|\stackrel{(a)}{=}  & | \GoodB_t (\Gc_T) |\stackrel{(b)}{\ge}  \nonumber \\
 | \GoodB_t (G(n, p_{\min}) )| \stackrel{(c)}{=} & | \Ac_t (G(n, p_{\min}))| \stackrel{(d)}{>}t. 
\end{align}
 In (\ref{neweq}),  equality (a) is an immediate consequence of  \eqref{neweq1},
inequality (b) holds by monotonicity of sets  $\GoodB_t $  with respect  to ``$\le_{\text{st}}$'',  
while   equality (c)  descends from Theorem~\ref{theorem:gross}.  
Inequality (d)  descends from the fact  that we assume $t<T_{G_{p_{\min}}}$.

Thus, necessarily,   $\Ac_T(\Gc_T)= T\ge \min(T_{G_{p_{\min}}}, T_{G_{p_{\max}}})= n- o(n)$ and $\BadB_T (\Gc_T) =\emptyset$.



\end{document}